\documentclass[aps,prb,10pt,twocolumn]{revtex4-1}
\usepackage{bm}
\usepackage{amsmath}
\usepackage{amsfonts}
\usepackage{amssymb}
\usepackage{float}
\usepackage[final]{graphicx}
\usepackage[caption = false]{subfig}
\usepackage{dsfont}
\usepackage[colorlinks=true,linktocpage=true,breaklinks=true]{hyperref}

\begin{document}
\title{Semi-realistic tight-binding model for spin-orbit torques}
\author{Guilhem Manchon$^1$}
\author{Sumit Ghosh$^1$}
\author{Cyrille Barreteau$^2$}
\author{Aur\'elien Manchon$^{1,3,4}$}
\email{manchon@cinam.univ-mrs.fr\\aurelien.manchon@kaust.edu.sa}
\affiliation{$^1$Physical Science and Engineering Division (PSE), King Abdullah University of Science and Technology (KAUST), Thuwal 23955-6900, Kingdom of Saudi Arabia}
\affiliation{$^2$SPEC, CEA, CNRS, Universit\'e Paris-Saclay, CEA Saclay, 91191 Gif-sur-Yvette, France}
\affiliation{$^3$Computer, Electrical and Mathematical Science and Engineering Division (CEMSE), King Abdullah University of Science and Technology (KAUST), Thuwal 23955-6900, Kingdom of Saudi Arabia}
\affiliation{$^4$Aix-Marseille Univ, CNRS, CINaM, Marseille, France}

\begin{abstract}
We compute the spin-orbit torque in a transition metal heterostructure using Slater-Koster parameterization in the two-center tight-binding approximation and accounting for d-orbitals only. In this method, the spin-orbit coupling is modeled within Russel-Saunders scheme, which enables us to treat interfacial and bulk spin-orbit transport on equal footing. The two components of the spin-orbit torque, dissipative (damping-like) and reactive (field-like), are computed within Kubo linear response theory. By systematically studying their thickness and angular dependence, we were able to accurately characterize these components beyond the traditional "inverse spin galvanic" and "spin Hall" effects. Whereas the conventional field-like torque is purely interfacial, we unambiguously demonstrate that the conventional the damping-like torque possesses both an interfacial and a bulk contribution. In addition, both field-like and damping-like torques display substantial angular dependence with strikingly different thickness behavior. While the planar contribution of the field-like torque decreases smoothly with the nonmagnetic metal thickness, the planar contribution of the damping-like torque increases dramatically with the nonmagnetic metal thickness. Finally, we investigate the self-torque exerted on the ferromagnet when the spin-orbit coupling of the nonmagnetic metal is turned off. Our results suggest that the spin accumulation that builds up inside the ferromagnet can be large enough to induce magnetic excitations.
\end{abstract}
\maketitle

\section{Introduction}
Current-driven spin-orbit torques have become a prodigal area of research in the past ten years \cite{Brataas2014,Manchon2019}. This magnetic torque enables the electrical control of ferromagnets using spin densities generated through angular momentum transfer between the orbital and spin degrees of freedom \cite{Miron2011b,Liu2012}. Understanding the physical origin of the torques, their symmetries and materials dependence has been the subject of intense collaborations between experimentalists and theorists. From the theory standpoint, several mechanisms have been identified, among which spin Hall effect \cite{Sinova2015}, inverse spin galvanic effect \cite{Manchon2008b,Manchon2009b,Garate2009} (also called Rashba-Edelstein effect), spin swapping \cite{Lifshits2009,Saidaoui2015b,Saidaoui2016}, interfacial spin precession\cite{Amin2018,Freimuth2018} etc. In spite of these efforts, important questions remain to be answered as experimental data point toward complex thickness, angular and temperature dependences \cite{Kim2013,Garello2013,Avci2014,Qiu2015,Ghosh2017}. Although initial oversimplified theories attributed the dissipative "damping-like" component of the torque (even under time reversal) to spin Hall effect\cite{Liu2011,Haney2013b} and the reactive "field-like" component (odd under time reversal) to the inverse spin galvanic effect\cite{Manchon2008b,Miron2010}, this crude picture has been severely questioned by the most recent experiments. It remains unclear whether the torque components can be solely attributed to spin Hall effect, inverse spin galvanic effect, or a combination of both. In addition, a recent series of experiments\cite{Fan2013,Baek2018,Safranski2019} have identified unexpected torque components that are attributed to mechanisms beyond spin Hall and inverse spin galvanic effects \cite{Saidaoui2016,Amin2018,Freimuth2018}. A detailed discussion on these open questions can be found in Ref. \onlinecite{Manchon2019}. 

In order to properly characterize and predict the behavior of spin-orbit torques in heterostructures, the model should be both comprehensive and transparent. As a matter of fact, such a model should ideally account for the realistic band structure of the heterostructure to treat bulk and interfacial spin-orbit effects on equal footing. But it should also be able to provide general trends that can serve as guidelines to experiments. To date, most models have addressed only certain aspects of the spin-orbit torques such as the interfacial inverse spin galvanic through model Hamiltonians \cite{Manchon2008b,Manchon2009b,Garate2009,Bijl2012,Li2015b,Qaiumzadeh2015,Ado2017} or spin Hall effect either through drift-diffusion or Boltzmann transport equation \cite{Haney2013b,Chen2017b}. Whereas these approaches are quite transparent, their main limitation is their inability to treat both interface and bulk effects altogether and in particular the neglect of interfacial orbital hybridization that is known to be crucial in transition metal multilayers\cite{Blugel2007,Grytsyuk2016,Wang2016b}. The common denominator between spin Hall and inverse spin galvanic effects is that they both stem from non-equilibrium orbital currents \cite{Tanaka2008,Jo2018} or densities \cite{Yoda2018} that involve specific admixture of atomic orbitals. Consequently, the proper modeling of spin-orbit effects in heterostructures beyond the Rashba and spin Hall phenomenologies requires a multi-orbital scheme.

To date, the most accurate approach to compute multi-orbital transport properties is to rely on density functional theory. This approach has been used extensively to compute spin and anomalous Hall effects in the bulk \cite{Yao2004,Guo2008,Lowitzer2011,Sun2016}, and has been recently extended to compute spin transport in heterostructures (e.g., Ref. \onlinecite{Haney2013a}). Various techniques have been proposed including Wannier interpolation of the band structure \cite{Freimuth2014a,Geranton2015,Geranton2016,Mahfouzi2018a}, Korringa-Kohn-Rostoker method \cite{Wimmer2016,Ebert2011b}, or real space Hamiltonian with tight-binding linear muffin-tin orbitals \cite{Wang2016,Belashchenko2019}. While the first two methods are well adapted to compute Kubo-Streda formula, the latter is suitable for two-terminal simulations, following the Landauer-B\"uttiker scheme. These different methods present the crucial advantage of modeling accurately the orbital hybridization across the whole structure. They are however computationally intensive, which makes them hardly adapted for systematic investigations (such as thickness dependence).\par

From this standpoint, developing a multi-orbital tight-binding model is an interesting option as it reduces the size of the matrices to deal with numerically\cite{Papaconstantopoulos2003,Papaconstantopoulos2015}. When interfaced with density functional theory, this method enables the accurate simulation of magnetic \cite{Barreteau2016} as well as transport properties \cite{Tanaka2008} in bulk materials. Using this approach, the intrinsic contribution to spin Hall effect has been computed in various materials (semiconductors\cite{Guo2005}, transition metals \cite{Yao2005,Tanaka2008,Freimuth2010}, topological insulators \cite{Sahin2015}, Weyl semimetals \cite{Sun2016} etc.) using the zero-temperature Berry curvature formula. Unfortunately, these results are only valid for vanishing disorder in the bulk, and cannot be transposed to experimentally relevant setup where injection through interface dominate \cite{Sinova2015}. To properly compute spin-charge conversion processes and spin-orbit torque, one needs to model the full heterostructure, including the interface \cite{Haney2013a,Freimuth2014a}. Recently, we applied this approach to a heterostructure made of a topological insulator capped with a (ferro- or antiferro-) magnetic material \cite{Ghosh2018,Ghosh2019}. In this work, each unit cell is modeled by a 4$\times$4 Hamiltonian matrix regularized on a cubic lattice \cite{Marchand2012}. Although quite crude, this approximation allowed us to model spin-orbit torque in various transport regimes and determine the minor role of spin Hall effect in these structures.

 In the present work, we use a multi-orbital tight-binding model to compute the spin-orbit torque in transition metal heterostructures. Whereas this method does not provide the accurate band structure obtained by density functional theory, it retains the most prominent features of the density of states, atomic spin-orbit coupling and interfacial orbital hybridization. It is also more flexible and computationally efficient, allowing for systematic characterization of the non-equilibrium properties of the heterostructure. In particular, we investigate the thickness and angular dependences of the torque components and obtain a large "planar" damping-like torque. We also investigate the nature of the self-torque, i.e. the spin-orbit torque taking place in the ferromagnet itself, and demonstrate that it can be substantial in spite of the large magnetic exchange \cite{Pauyac2018,Wang2019b}.

\section{Model and formalism\label{s:model}}

In this section, we first introduce a toy model to discuss how the interfacial orbital mixing gives rise to "Rashba-like" spin-orbit coupling. Then, we describe the tight-binding model of the heterostructure, and finally we expose the formalism we use to compute the transport properties.

\subsection{Interfacial spin splitting with p and d orbitals\label{s:porbitals}}
In centrosymmetric materials, such as the transition metals we consider in this work, the spin Hall effect occurring in the bulk is usually attributed to intrinsic origin \cite{Murakami2003,Sinova2004}, i.e. to the Berry curvature of the wave functions. Following the scenario established by Tanaka et al. \cite{Tanaka2008,Kontani2009,Jo2018}, Berry curvature in momentum space creates an orbital Hall current, which is spin-polarized by turning on the atomic spin-orbit coupling. In contrast, little is known about the orbital origin of the interfacial "Rashba" spin-orbit coupling. Since the early works on this topic \cite{Vasko1979,Ohkawa1974,Bychkov1984}, it was proposed that upon inversion symmetry breaking, the spin-orbit coupling experienced by the Bloch electrons acquires a momentum-dependent Zeeman energy term, usually written
\begin{eqnarray}
{\cal H}_{\rm R}=\alpha_{\rm R}\hat{\bm\sigma}\cdot(\hat{\bf p}\times {\bf z}),
\end{eqnarray}
where $\alpha_{\rm R}$ is called the Rashba parameter. In their pioneering work, Petersen and Hedeg\aa rd \cite{Petersen2000} considered the Rashba spin splitting of Au (111) surface and proposed that the surface potential facilitates the admixture between p$_z$ and p$_{x,y}$ orbitals. This hybridization results in Rashba spin-orbit coupling when atomic spin-orbit coupling is turned on. A similar idea was put forward by Bihlmayer et al. \cite{Bihlmayer2006}, suggesting that inversion symmetry breaking promotes the admixture between $l$ and $l\pm1$ orbitals. In this section, we wish to provide an explicit derivation of this effect and establish a direct connection between orbital mixture due to inversion symmetry breaking and Rashba-like spin-orbit coupling.

\begin{figure}
\begin{center}
        \includegraphics[width=8cm]{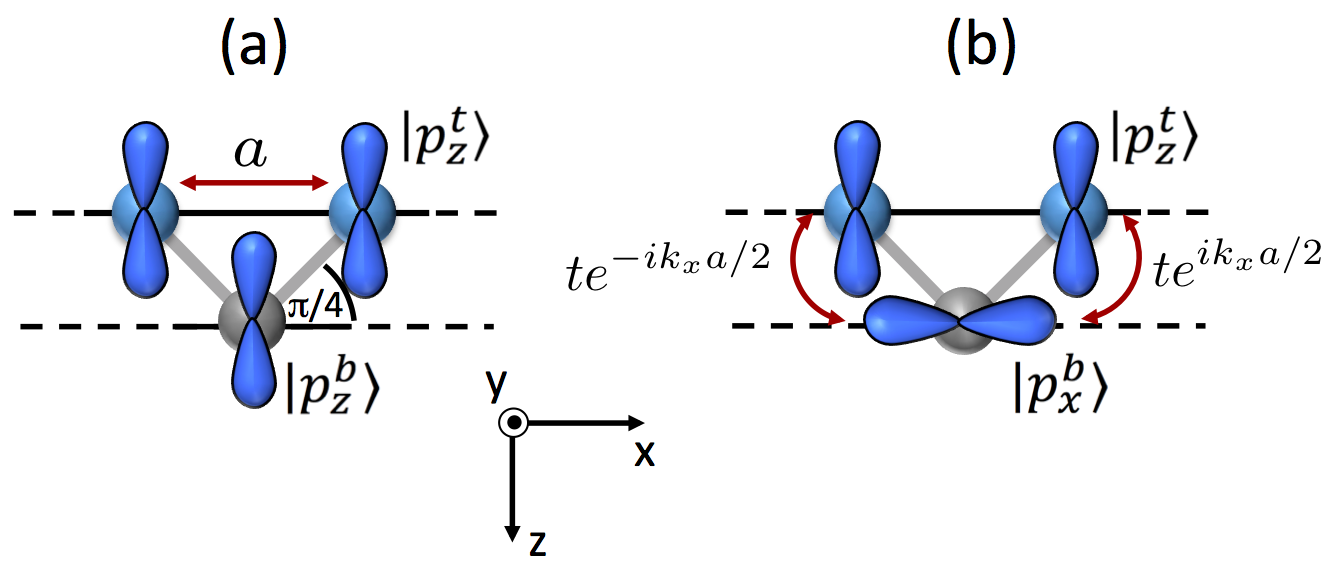}
      \caption{(Color online) Schematics of the diatomic chain model. The atoms of the bottom chain (gray) possess both p$_z$ (a) and p$_x$ (b) orbitals, while the atoms of the top chain has only p$_z$ orbitals. The phase acquired by Bloch electrons hopping from one orbital to the other is also given.\label{FigRashbap}}
\end{center}
\end{figure}

Let us consider a chain of atoms, extended along $x$ and with all three p$_x$, p$_y$ and p$_z$ orbitals. We can discard the p$_y$ orbitals from our discussion  right away since they don't couple to either p$_x$, or p$_z$. We now break the inversion symmetry by coupling this chain with another chain of atoms with only p$_z$ orbitals. The system is depicted on Fig. \ref{FigRashbap}(a) and (b). The atoms of the bottom chain (with both p$_x$ and p$_z$) are represented in gray and the atoms of the top chain (with p$_z$ only) are in light blue. In the two-center tight-binding approximation and the $\{{\rm p}_z^t,{\rm p}_z^b,{\rm p}_x^b\}$ basis, the Hamiltonian of this diatomic chain reads

\begin{equation}
{\cal H}_{\rm chain}=
\left(\begin{matrix}
\varepsilon_{k}^{t}& V_{zz} &V_{zx} \\
V_{zz} &\varepsilon_{k}^{z}&0\\
V_{zx}^*  &0&\varepsilon_{k}^{x}\\
\end{matrix}\right).
\end{equation}
Here p$_\nu^\eta$ refers to the $\nu$-th orbital of the top ($\eta=t$) or bottom chain ($\eta=b$), $V_{zz}=(V_\sigma+V_\pi)\cos k_xa/2$ and $V_{zx}=-i(V_\sigma-V_\pi)\sin k_xa/2$, $V_{\sigma,\pi}$ being the Slater-Koster hopping integrals \cite{Slater1954}. In order to keep our result analytically tractable, we assume that $\varepsilon_{k}^{z}=\varepsilon_{k}^{x}$. Then, we end up with three bands with dispersion,
\begin{eqnarray}
\varepsilon_{\bf k}^0&=&\varepsilon_{k}^{ z},\;\varepsilon_{\bf k}^\pm=\frac{\varepsilon_{k}^{t}+\varepsilon_{k}^{z}}{2}\pm\frac{1}{2}\Delta_k,
\end{eqnarray}
with $\Delta_k=\sqrt{(\varepsilon_{k}^{t}-\varepsilon_{k}^{z})^2+4(|V_{zz}|^2+|V_{zx}|^2)}$. The corresponding eigenstates read
\begin{eqnarray}
|0\rangle&=&\frac{1}{\sqrt{|V_{zz}|^2+|V_{zx}|^2}}\left(-V_{zx}|{\rm p}^b_z\rangle+V_{zz}|{\rm p}^b_x\rangle\right),\\
|+\rangle&=&\cos\chi|{\rm p}_z^t\rangle+\frac{\sin\chi}{\sqrt{|V_{zz}|^2+|V_{zx}|^2}}\left(V_{zz}|{\rm p}_z^b\rangle+V_{zx}^*|{\rm p}_x^b\rangle\right),\\
|-\rangle&=&-\sin\chi|{\rm p}_z^t\rangle+\frac{\cos\chi}{\sqrt{|V_{zz}|^2+|V_{zx}|^2}}\left(V_{zz}|{\rm p}_z^b\rangle+V_{zx}^*|{\rm p}_x^b\rangle\right),\nonumber\\
\end{eqnarray}
where $\cos2\chi=(\varepsilon_{k}^t-\varepsilon_{k}^z)/\Delta_k$. We now evaluate the orbital momentum on the bottom chain, and using $\langle {\rm p}^b_x|{\bf L}|{\rm p}^b_z\rangle=i{\bf y}$, we get
\begin{eqnarray}
\langle 0|{\bf L}|0\rangle&=&-\frac{2{\rm Im}\left[V_{zz}V_{zx}^*\right]}{\sqrt{|V_{zz}|^2+|V_{zx}|^2}}{\bf y},\\
\langle +|{\bf L}|+\rangle&=&\sin^2\chi\frac{2{\rm Im}\left[V_{zz}V_{zx}^*\right]}{\sqrt{|V_{zz}|^2+|V_{zx}|^2}}{\bf y},\\
\langle -|{\bf L}|-\rangle&=&\cos^2\chi\frac{2{\rm Im}\left[V_{zz}V_{zx}^*\right]}{\sqrt{|V_{zz}|^2+|V_{zx}|^2}}{\bf y},
\end{eqnarray}
where $2{\rm Im}\left[V_{zz}V_{zx}^*\right]=[(V_\sigma)^2-(V_\pi)^2]\sin k_xa$. This toy model shows that, due to the lack of inversion symmetry, the eigenstates of the diatomic chain acquire an orbital momentum that is {\em odd} in linear momentum $k$. This orbit-momentum locking results in orbital Edelstein effect \cite{Yoda2018}, i.e. the electrical generation of an orbital magnetic moment. \par

When atomic spin-orbit coupling is turned on, the spin momentum of the Bloch electron aligns on its orbital momentum. Therefore, in the $\{|0\rangle,|+\rangle,|-\rangle\}$ basis, the spin-diagonal Hamiltonian of these Bloch states acquires an off-diagonal contribution,
\begin{eqnarray}\label{eq:rashba}
\langle\xi_{\rm so}{\bf L}\cdot\hat{\bm \sigma}\rangle=
{\cal M}_{\rm so}\frac{[(V_\sigma)^2-(V_\pi)^2]\sin k_xa}{\sqrt{[V_\sigma)^2+(V_\pi)^2+2V_\sigma V_\pi \cos k_xa}}\hat{\sigma}_y,\nonumber\\
\end{eqnarray}
where $\hat{\bm \sigma}$ is the vector of Pauli spin matrices, and ${\cal M}_{\rm so}={\rm Diag}(-\xi_{\rm so}^t,\xi_{\rm so}^b\sin^2\chi,\xi_{\rm so}^b\cos^2\chi)$, Diag(...) being the diagonal matrix and $\xi_{\rm so}^\eta$ the spin-orbit coupling energy of the $\eta$-th chain ($\eta=t,b$). This Hamiltonian explicitly connects the linear momentum $k_x$ with the spin momentum $\sigma_y$, resulting in Rashba and Dzyaloshinskii-Moriya effects \cite{Manchon2015}. This model can be straightforwardly extended to higher dimensions and higher order orbitals (d, f etc.). In the case of a transition metal interface, the orbital admixture required to obtain a spin density along $S_y$ for an electron propagating along $x$ is typically d$_{xy}$-d$_{yz}$, d$_{zx}$-d$_{z^2}$ or d$_{zx}$-d$_{x^2-y^2}$. A microscopic model of spin-orbit effects at interfaces should {\em a minima} contain these orbitals.

\subsection{Tight-binding model of the heterostructure}

\begin{figure}
        \includegraphics[width=8cm]{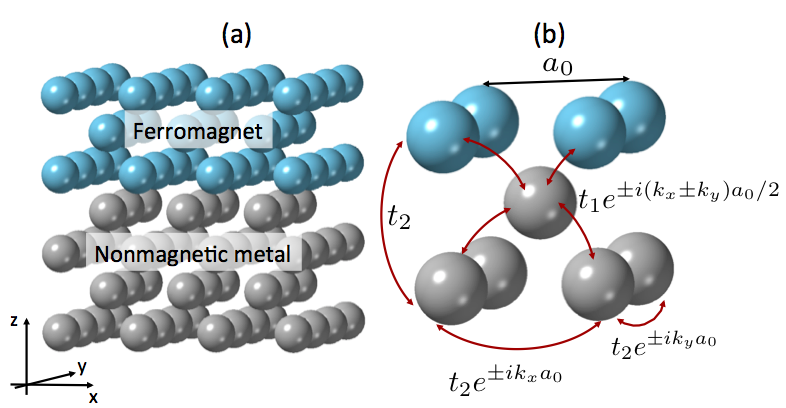}
      \caption{(Color online) (a) Schematics of the bcc heterostructure composed of a ferromagnetic metal (blue) deposited on top of a nonmagnetic metal (gray). For simplicity, we consider that both metals possess the same lattice parameter. (b) Hopping parameters at the interface between the ferromagnet and the nonmagnetic metal. $t_1$ stands for the nearest neighbor hopping and $t_2$ stands for the second nearest neighbor hopping.\label{Fig0}}
\end{figure}

We now move on to the description of the tight-binding model of our transition metal heterostructure, depicted on Fig. \ref{Fig0}(a). This heterostructure consists of two adjacent metallic slabs with bcc crystal structure along the (001) direction, possessing the same lattice parameter. Each metallic slab is constituted of monolayers stacked on top of each other. Considering the ten d-orbitals, each monolayer adopts a square lattice described by the Hamiltonian

\begin{equation}
{\cal H}_{\rm mono}=
\left(\begin{matrix}
\gamma_{xy}^{\bf k}&0 &0  & t_{xy,z^2}^{\bf k} &0\\
0&\gamma_{yz}^{\bf k}&t_{zx,yz}^{{\bf k}} &0 & 0\\
0  &t_{zx,yz}^{{\bf k},*}&\gamma_{zx}^{\bf k}&0 &0\\
t_{xy,z^2}^{{\bf k},*} &0&0&\gamma_{z^2}^{\bf k}&0\\
0&0&0&0&\gamma_{x^2-y^2}^{\bf k}\\
\end{matrix}\right)\label{eq:Hmono}
\end{equation}
where the parameters $\gamma_{\nu}^{\bf k}$ and $t_{\mu,\nu}^{\bf k}$ are given explicitly in the Appendix. This Hamiltonian is written in the basis $\{{\rm d}_{xy},{\rm d}_{yz},{\rm d}_{zx},{\rm d}_{z^2},{\rm d}_{x^2-y^2}\}$. This form is valid for each spin species, so that the spin-dependent Hamiltonian reads ${\cal H}_{\rm mono}\otimes\hat{\sigma}_0$. In addition, we define the exchange Hamiltonian, ${\cal H}_{\rm ex}$, as
\begin{eqnarray}
{\cal H}_{\rm ex}=\frac{1}{2}{\rm Diag}\left(\Delta_{xy},\Delta_{yz},\Delta_{zx},\Delta_{z^2},\Delta_{x^2-y^2}\right)\otimes\hat{\bm\sigma}\cdot{\bf m}.\nonumber\\
\end{eqnarray}
Here $\Delta_{\nu}$ is the exchange energy of the $\nu$-th d orbital and ${\bf m}$ is the ferromagnetic order parameter. Hence, the Hamiltonian for a square lattice monolayer is a 10$\times$10 matrix. In addition, one needs to account for the spin-orbit coupling matrix,
\begin{equation}
{\cal H}_{\rm soc}=\xi_{\rm so}
\left(\begin{matrix}
0&i\hat{\sigma}_y &-i\hat{\sigma}_x  &0&2i\hat{\sigma}_z\\
-i\hat{\sigma}_y&0&i\hat{\sigma}_z &-i\sqrt{3}\hat{\sigma}_x &-i\hat{\sigma}_x\\
i\hat{\sigma}_x  &-i\hat{\sigma}_z&0&i\sqrt{3}\hat{\sigma}_y &-i\hat{\sigma}_y\\
0&i\sqrt{3}\hat{\sigma}_x&-i\sqrt{3}\hat{\sigma}_y &0&0\\
-2i\hat{\sigma}_z&i\hat{\sigma}_x&i\hat{\sigma}_y&0&0\\
\end{matrix}\right).\label{eq:soc}
\end{equation}
The Hamiltonian of a monolayer is therefore
\begin{equation}
{\cal H}_{0}={\cal H}_{\rm mono}\otimes\hat{\sigma}_0+{\cal H}_{\rm ex}+{\cal H}_{\rm soc}.
\end{equation}
This monolayer is connected to the top nearest monolayer by the matrix ${\cal T}_1$ whose elements are the nearest neighbor hopping parameters $t_{\mu,\nu}^{z,{\bf k}}$ between orbital d$_{\mu}$ in the bottom layer and orbital d$_\nu$ in the top layer. The connection to the second top nearest monolayer is accounted for by the matrix ${\cal T}_2$. The elements of both matrices are given explicitly in the Appendix. The Hamiltonian of one bcc slab is then defined
\begin{equation}\label{Hlayer}
\cal{H}_{\rm layer}=
\left(\begin{matrix}
{\cal H}_0 & {\cal T}_1 & {\cal T}_2 & 0&\\
{\cal T}^\dagger_1 & {\cal H}_0& {\cal T}_1 & {\cal T}_2  &\ddots\\
{\cal T}^\dagger_2 &{\cal T}^\dagger_1& {\cal H}_0&{\cal T}_1&\ddots \\
0&{\cal T}^\dagger_2 &{\cal T}^\dagger_1&{\cal H}_0&\ddots\\
& \ddots&\ddots& \ddots&\ddots\\
\end{matrix}\right).
\end{equation}

\begin{table}
\begin{tabular}{c|cccccc}
& $V_\sigma^1$ & $V_\pi^1$ & $V_\delta^1$& $V_\sigma^{2}$ & $V_\pi^2$ & $V_\delta^2$\\\hline
FM & -0.618 & 0.37 &-0.035&-0.37&0.08 & 0.01\\
NM & -1.61 & 0.71 &0.034&-0.99 & -0.17 & 0.12\\
FM/NM & -1.11 & 0.54 &-0.001 & -0.68 & -0.046 &0.066\\
\end{tabular}
\begin{tabular}{c|ccccc}
&$\varepsilon_{xy,yz,zx}$&$\varepsilon_{z^2,x^2-y^2}$&$\Delta_{xy,yz,zx}$&$\Delta_{z^2,x^2-y^2}$&$\xi_{\rm soc}$\\\hline
FM & 12.8 & 12.5 &1.85&1.73&0.065 \\
NM & 13.16 & 11.66 & 0 &0 & 0.367  \\
\end{tabular}
\caption{Slater-Koster parameters used to model the FM/NM heterostructure and extracted from Ref. \onlinecite{Papaconstantopoulos2015}. The parameters are given in eV.\label{table1}}
\end{table} 

The matrix elements of ${\cal H}_{\rm mono}$, ${\cal T}_1$ and ${\cal T}_2$ are written in terms of the two-site Slater-Koster parameters (see Appendix) given for each slab in Table \ref{table1}. We adopt the parameters computed by Papaconstantopoulos \cite{Papaconstantopoulos2015} for {\em bulk} bcc Fe and bcc W. The lattice parameter of both slabs is set to that of bulk bcc W, $a_0=3.155$ \AA, imposing 9\% lattice mismatch with bcc Fe whose bulk lattice parameter is 2.866 \AA. Notice that the onsite energies of the ferromagnetic orbitals are rigidly shifted by an offset $\varepsilon_0$ compared to their value in bulk Fe in order to allow for band structure alignement between the ferromagnetic and nonmagnetic metals (see below). With these parameters, we determine the Hamiltonian for the nonmagnetic (NM) and ferromagnetic (FM) slabs, $\cal{H}_{\rm NM}$ and $\cal{H}_{\rm F}$. \par

Finally, the heterostructure is obtained by stitching the two individual slabs together.
\begin{equation}
\cal{H}=
\left(\begin{matrix}
{\cal H}_{\rm F} & {\cal T}_{\rm FN} \\
{\cal T}_{\rm FN}^\dagger & {\cal H}_{\rm NM}\\
\end{matrix}\right)
\end{equation}
The hopping matrix ${\cal T}_{\rm FN}$ is simply given by ${\cal T}_1$ and ${\cal T}_2$ adopting the parameters of Table \ref{table1}. In the absence of further knowledge, the hopping parameters between the highest nonmagnetic layer and the lowest magnetic layer are taken as the average of the bulk hopping parameters of Fe and W. Ideally, one would need to fit the tight-binding parameters to the band structure of the heterostructure computed self-consistently from first principles \cite{Barreteau2016}, which remains out of the scope of the present work but constitutes an appealing development of the present work. Indeed, we emphasize that the nonmagnetic transition metal is expected to acquire interfacial magnetization by proximity with the ferromagnetic metal \cite{Grytsyuk2016}. This induced magnetization is neglected in our model because our tight-binding parameters are that of the bulk materials. Nevertheless, a previous first principles investigation of the Pt/Co(111) interface has shown that such an induced magnetization has minor effect on the spin-orbit torque (see Fig. 7 in Ref. \onlinecite{Haney2013a}).\par
\begin{figure}
        \includegraphics[width=6cm]{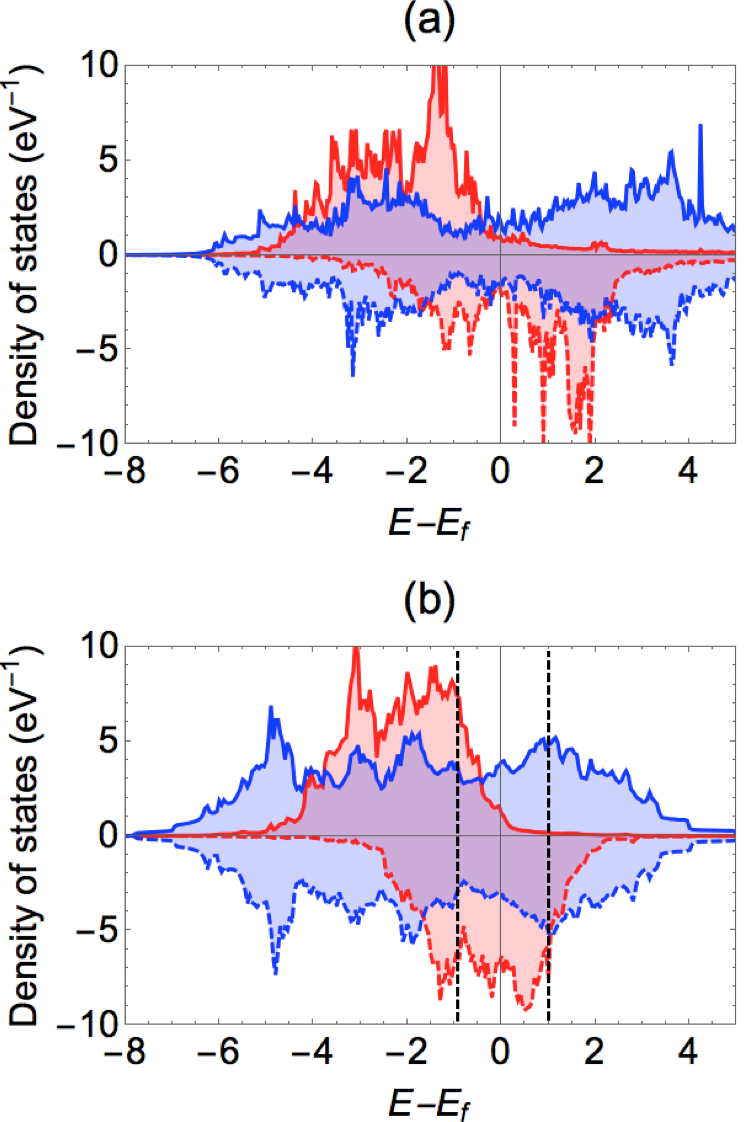}
      \caption{(Color online) Spin-resolved density of states of Fe(5)/W(7) projected on the d-orbitals and calculated by (a) density functional theory and (b) our tight-binding model (with $\Gamma=10$ meV). The blue shaded region corresponds to W (nonmagnetic layer) while the red shaded region corresponds to Fe (magnetic layer). The vertical dashed lines in (b) correspond to two cases of interest discussed in Section \ref{s:prof}. \label{Fig1}}
\end{figure}

Considering the numerous approximations we took (first and second nearest neighbor hopping only, no self-consistent computation of the interfacial and exchange potentials, constrained lattice parameter, neglect of s and p orbitals, etc.), we do not expect our tight-binding model to accurately represent a realistic Fe/W bilayer. Nonetheless, we benchmarked our tight-binding model against the density of states of a Fe/W bilayer computed by density functional theory in order to enforce its reliability. These simulations have been conducted using Vienna ab initio simulation package (VASP) \cite{Kresse1996a,Kresse1996b} with PAW-PBE GGA pseudopotentials \cite{Blochl1994, Kresse1999}. The structure has been relaxed until forces on all the atoms go below 0.001 eV/$\rm \AA$ allowing both the atomic coordinates and lattice vectors to change. We have used  an energy cut off of 500 eV. For self consistent cycles we have used a $16 \times 16 \times 1$ k-mesh and for the density of states, we have used a $24 \times 24 \times 1$ k-mesh. We have neglected the effect of spin-orbit coupling and conducted a spin-polarized calculation as we are interested in spin-resolved density of states. Including spin-orbit coupling does not make any drastic change in the total density of states.\par

The first principles density of states of Fe(5)/W(7) projected on the d-orbitals only is reported on Fig. \ref{Fig1}(a) together with the density of states obtained for our FM(5)/NM(7) system [Fig. \ref{Fig1}(b)]. The figures in parenthesis indicate the number of monolayers, and the density of states is defined $-\frac{1}{\pi}{\rm Im}[\hat{G}^R]$, where $\hat{G}^R=(\varepsilon-{\cal H}+i\Gamma)$ is the retarded Green's function and $\Gamma$ is the homogeneous broadening. The tight-binding density of state in Fig. \ref{Fig1}(b) is obtained for a rigid energy shift $\varepsilon_0=3.1$ eV, and the Fermi energy is fixed at 14 eV in order to qualitatively reproduce the balance between up and down Fermi electrons obtained by VASP. With these parameters, the total number of electrons in the ferromagnetic and nonmagnetic metals are $n^{\rm FM}_\uparrow=4.6$, $n^{\rm FM}_\downarrow=2.34$ and $n^{\rm NM}_\uparrow+n^{\rm NM}_\downarrow=4.73$.\par

We immediately observe a number of differences between the two densities of states in terms of bandwidth and peak position. These differences are attributed to the crude approximations of the tight-binding model mentioned above. Nevertheless, both densities of states display the same essential features: similar bandwidth, spin splitting of the ferromagnetic metal, large overlap between the two materials close to Fermi level etc. Therefore, although our FM/NM heterostructure does not reproduce the ideal Fe/W case, it is a good representative of transition metal heterostructures.\par

\begin{figure}
\subfloat{\includegraphics[width = 4.9cm]{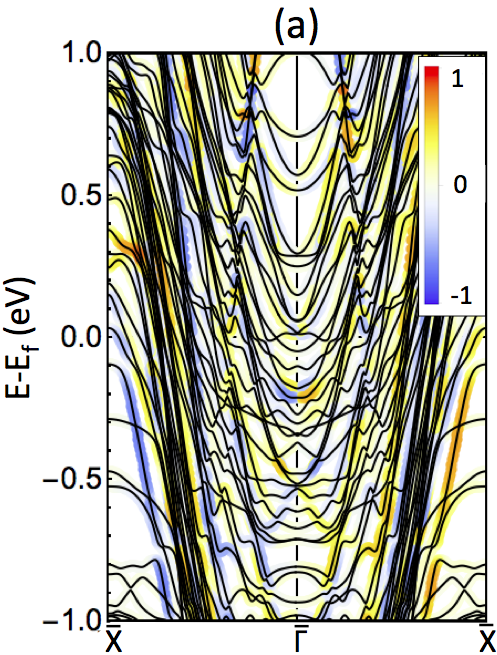}}\\
\subfloat{\includegraphics[width = 4.9cm]{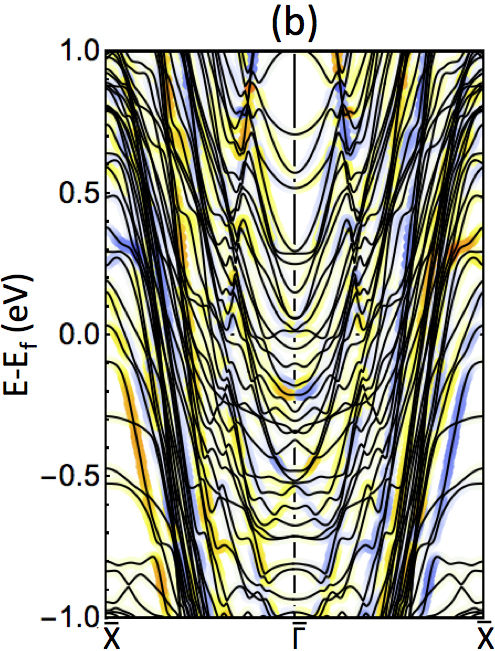}}\\
\subfloat{\includegraphics[width = 4.9cm]{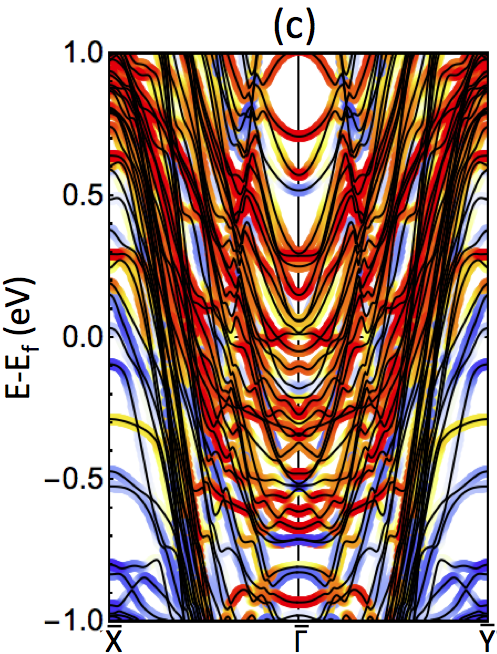}}
\caption{(Color online) Spin-resolved band structure for FM(5)/NM(7) bilayer: (a) $s_x$ along $\bar{\rm Y}-\bar{\Gamma}-\bar{\rm Y}$, (b) $s_y$ along $\bar{\rm X}-\bar{\Gamma}-\bar{\rm X}$, and (c) $s_z$ along $\bar{\rm X}-\bar{\Gamma}-\bar{\rm Y}$. Blue and red colors refer to opposite sign of the spin momentum.}
\label{FigBS}
\end{figure}

We conclude this discussion by considering the spin texture in momentum space. As explained above, symmetry breaking at the interface results in orbital Edelstein effect, which promotes the onset of spin-momentum locking in the presence of spin-orbit coupling. Figure \ref{FigBS} shows the band structure around $\bar{\Gamma}$ point projected on the spin momentum components, $s_{x,y,z}=\langle\hat{\sigma}_{x,y,z}\rangle $. In this calculation, the magnetization is set along ${\bf z}$. Figure \ref{FigBS}(a) displays $s_x$ component when spanning the momentum between $\bar{\rm Y}$ and $\bar{\Gamma}$ points, Fig. \ref{FigBS}(b) displays $s_y$ component when spanning the momentum between $\bar{\rm X}$ and $\bar{\Gamma}$ points, and Fig. \ref{FigBS}(c) displays $s_z$ component along the $\bar{\rm X}-\bar{\Gamma}-\bar{\rm Y}$ path. The in-plane spin texture is antisymmetric in momentum and displays the $\hat{\bm\sigma}\sim {\bf z}\times{\bf k}$ symmetry expected for Rashba spin-orbit coupling. In contrast, the $s_z$ component is symmetric and reflects the spin polarization of the bands due to magnetic exchange.

\subsection{Transport formalism}

The transport properties are computed using Kubo-Streda formula \cite{Sinitsyn2006,Freimuth2014a}. In this framework, the conductivity tensor reads
\begin{eqnarray}
\sigma_{ij}&=&\frac{e\hbar}{2\pi}\int d\varepsilon \partial_\varepsilon f(\varepsilon){\rm Tr}\left[\hat{v}_j\hat{G}^R\hat{v}_i(\hat{G}^R-\hat{G}^A)\right].
\end{eqnarray}
Here, ${\rm Tr}$ denotes the trace over the orbital, spin and monolayer degrees of freedom and the sum over the Brillouin zone, $e=-|e|$ is the electron charge, and ${\hat v}_i=(1/\hbar)\partial_{{\bf k}_i}{\cal H}$ is the velocity operator. The local spin density on monolayer $\eta$ per unit electric field reads
\begin{equation}
{\bf S}_\eta=\frac{e\hbar}{2\pi}\int d\varepsilon \partial_\varepsilon f(\varepsilon){\rm Tr}\left[{\hat P}_\eta\otimes\hat{\bm\sigma}\hat{G}^R\hat{v}_i(\hat{G}^R-\hat{G}^A)\right],
\end{equation} 
where ${\hat P}_\eta$ is the projector on monolayer $\eta$. By construction, the matrix elements of ${\hat P}_\eta$ are equal to the 5$\times$5 identity matrix $\mathds{I}_{5}$ at the position of layer $\eta$ and zero elsewhere,
\begin{equation}
{\hat P}_\eta=
\left(\begin{matrix}
\ddots && &&\\
 &0& && \\
& &\mathds{I}_{5}&&\\
& &&0&\\
& && &\ddots\\
\end{matrix}\right).
\end{equation} 
 The torque per unit electric field is defined as
 \begin{equation}
{\bf T}=-\frac{e\hbar}{2\pi}\int d\varepsilon \partial_\varepsilon f(\varepsilon){\rm Tr}\left[{\bf m}\times{\bm \Omega}_{\rm ex}\hat{G}^R\hat{v}_i(\hat{G}^R-\hat{G}^A)\right]
\end{equation} 
where ${\bf m}\times{\bm \Omega}_{\rm ex}=-{\bf m}\times\partial_{\bf m}{\cal H}$ is the torque operator. In the remaining of the article, the conductivity of the slab is defined as $\sigma_{ij}/t$, $t$ being the thickness of the full heterostructure. The local spin density per unit electric field is in $m^{-1}$ and the torque is expressed as a spin conductivity, in the units of $(\hbar/2e)~\Omega^{-1}\cdot m^{-1}$. The disorder is accounted for through a homogeneous broadening $\Gamma$. Under this approximation, no higher order scattering events are taken into account (e.g., skew scattering, spin swapping etc.).

\section{Currents-driven spin-orbit torques in FM/NM heterostructure}
\subsection{Spin density profile\label{s:prof}}
\begin{figure}
        \includegraphics[width=6cm]{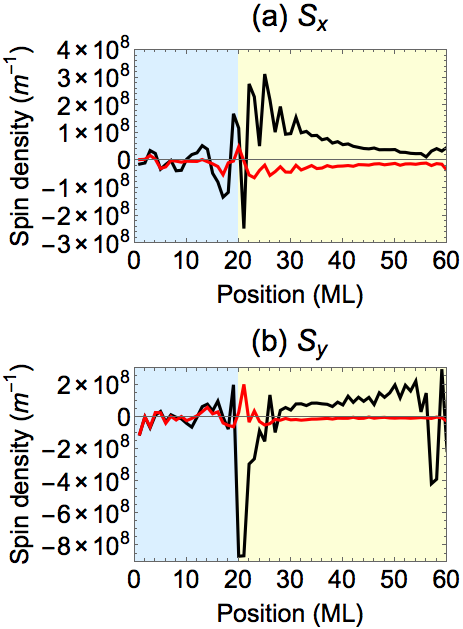}
      \caption{(Color online) Non-equilibrium in-plane spin density profile per unit electric field across the FM/NM bilayer. The shaded blue area refers to the FM region and the shaded yellow area refers to the NM region. The black solid line is the spin density when the spin-orbit coupling of both ferromagnetic and nonmagnetic layers is turned on, and the red solid line is the spin density when only the spin-orbit coupling of the ferromagnetic layer is on. Here the magnetization points perpendicular to the plane, along ${\bf z}$.\label{Fig2}}
\end{figure}

We first compute the current-driven spin density profile throughout the heterostructure, when the magnetization ${\bf m}$ points out of plane (${\bf m}\|{\bf z}$). The two in-plane components, $S_x$ and $S_y$, are given in Figs. \ref{Fig2}(a) and (b), respectively. The black curves correspond to the case where spin-orbit coupling is present in both ferromagnetic and nonmagnetic layers, while the red curves correspond to the case where only the ferromagnetic layer possesses spin-orbit coupling (see Section \ref{s:self}). When spin-orbit coupling is present in both ferromagnetic and nonmagnetic layers, we observe a clear accumulation of $S_x$ and $S_y$ components in the nonmagnetic metal. It is instructive to notice that the scale over which the spin density accumulates close to the interface is different for the two components. The $S_y$ component is localized close to the interface and vanishes quickly over about 10 monolayers (ML - corresponding to about 1.3 nm), while $S_x$ slowly decays over a few tens of ML (i.e., about 5 nm). Notice also that $S_x$ penetrates deeper in the ferromagnetic layer than $S_y$. This distinction suggests that $S_x$ is controlled by non-local transport processes (e.g., scattering and diffusion), while $S_y$ is much more localized at the interface. Finally, a last important feature that distinguishes $S_x$ and $S_y$ is the presence of a non-vanishing $S_y$ component close to the outer surface of the nonmagnetic layer. These two features are consistent with the standard representation of spin-orbit torque as arising from diffusive spin Hall effect and interfacial Rashba-like effect. As a result, one expects the torque to display two components, conventionally referred to as field-like and damping-like components and reading
\begin{eqnarray}\label{eq:fl}
{\bf T}_{\rm FL}&=&\tau_{\rm FL}{\bf m}\times({\bf z}\times{\bf E}),\\
{\bf T}_{\rm DL}&=&\tau_{\rm DL}{\bf m}\times[({\bf z}\times{\bf E})\times{\bf m}].\label{eq:dl}
\end{eqnarray}

\begin{figure}
        \includegraphics[width=7cm]{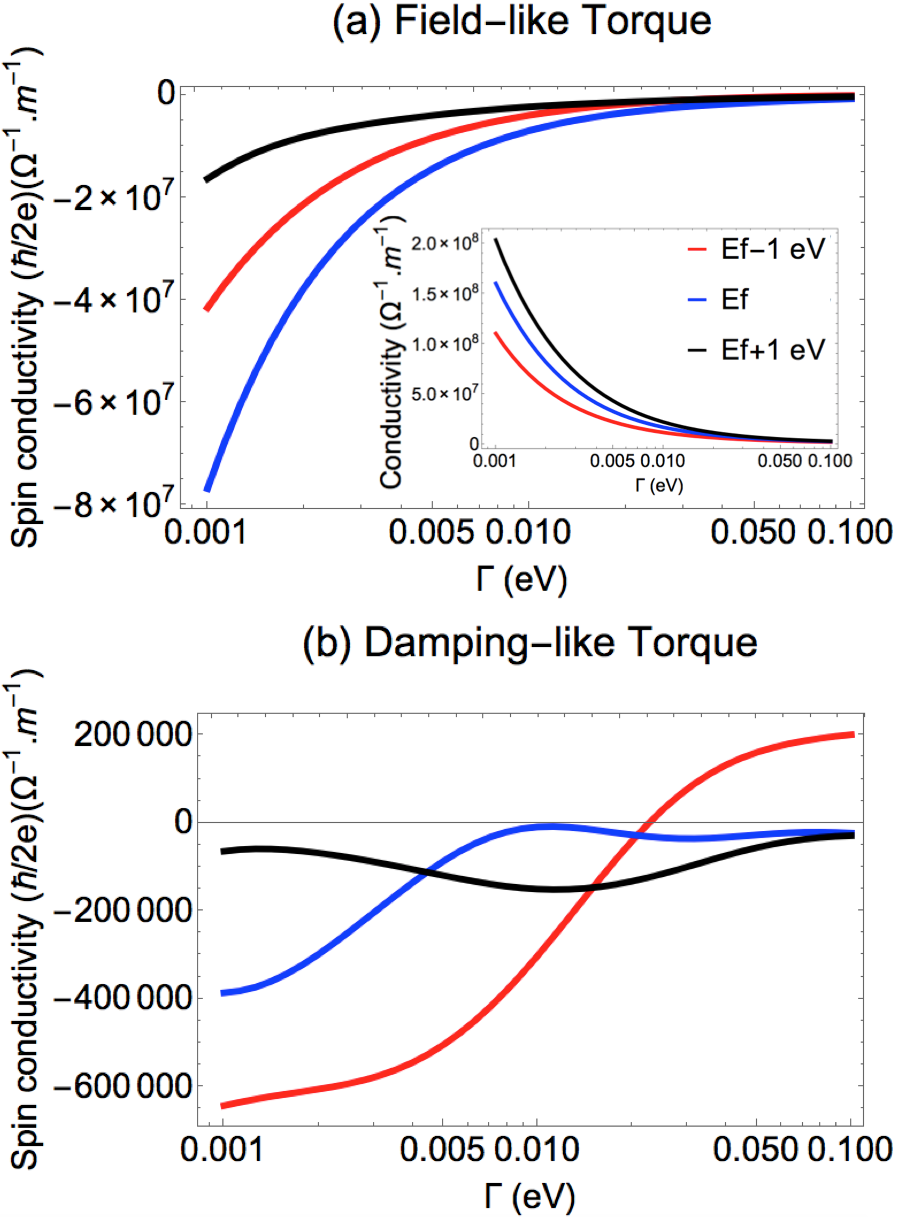}
      \caption{(Color online) Dependence of the two torque components, (a) field-like torque and (b) damping-like torque, as a function of the homogeneous broadening $\Gamma$, for different values of the transport energy, $E-E_{\rm f}=1$ eV (black), $E-E_{\rm f}=0$ eV (blue) and $E-E_{\rm f}=-1$ eV (red). The inset displays the slab conductivity. Here the magnetization points perpendicular to the plane, along ${\bf z}$.\label{Fig3}}
\end{figure}

We conclude this preliminary study by computing the torque exerted on the ferromagnetic layer as a function of the disorder, shown in Fig. \ref{Fig3}. The disorder-dependence of the torque components has been extensively used in previous studies to identify their physical origin \cite{Freimuth2014a,Li2015b}: a $1/\Gamma$-dependence, resembling the one of conductivity, suggests that extrinsic, intraband-dominated processes are involved, while a constant value when $\Gamma\rightarrow0$ indicates that intrinsic, interband-dominated processes govern the effect. Figure \ref{Fig3} displays the disorder-dependence of the (a) field-like and (b) damping-like components for three different Fermi energies, corresponding to different hybridization conditions as indicated by the dashed vertical lines in Fig. \ref{Fig1}(b). The conductivity and field-like torque both show $1/\Gamma$-dependence, confirming the intraband and extrinsic origin of this component (see, e.g., Ref. \onlinecite{Li2015b}). The damping-like torque saturates for $\Gamma\rightarrow0$, as expected for an interband intrinsic effect, but shows a more irregular behavior and even a change of sign for large disorder strength. In summary, the disorder dependence computed in Fig. \ref{Fig3} is consistent with the previous calculations of spin-orbit torque, both assuming a model Hamiltonian \cite{Li2015b} and using realistic density functional theory \cite{Freimuth2014a}.

\subsection{Thickness dependence\label{s:thickn}}

We now address the thickness dependence of the two torque components, a property that has been investigated in numerous experiments \cite{Kim2013,Fan2014c,Pai2015,Skinner2014,Nguyen2016,Ghosh2017}. To the best of our knowledge, such a thickness dependence has not been computed within density functional theory due to the prohibitive numerical cost. Hence, it has only been addressed using phenomenological models based on drift-diffusion or Boltzmann transport equations \cite{Manchon2012,Haney2013b,Amin2016b,Fischer2016}. In these works, the inverse spin galvanic effect is modeled by an interfacial Rashba interaction and the spin Hall effect is modeled using bulk drift-diffusion (e.g., Refs. \onlinecite{Shchelushkin2005b,Pauyac2018}). These models disregard quantum and semiclassical size effects as well as higher order scattering events such as spin swapping \cite{Saidaoui2015b,Saidaoui2016} and interfacial spin precession \cite{Amin2018}. The only physical mechanism giving rise to a non-trivial thickness dependence within these approaches is the spin relaxation in the nonmagnetic layer. In this context, the magnitude of both torque components follows a $\sim 1-\cosh^{-1}(t_{\rm NM}/\lambda_{\rm sf})$ law, where $t_{\rm NM}$ is the nonmagnetic layer thickness and $\lambda_{\rm sf}$ is its spin relaxation length. This law has been confirmed, at least phenomenologically, in several experimental studies \cite{Kim2013,Hayashi2014} (see also Fig. 24 in Ref. \onlinecite{Manchon2019}). However, at very small thicknesses ($\approx0.5$ nm for Ta substrate and $\approx2$ nm for Hf substrate), a change of sign of the torque components has been reported that remains unexplained \cite{Kim2013,Akyol2016,Ramaswamy2016}.\par

\begin{figure}
        \includegraphics[width=8.5cm]{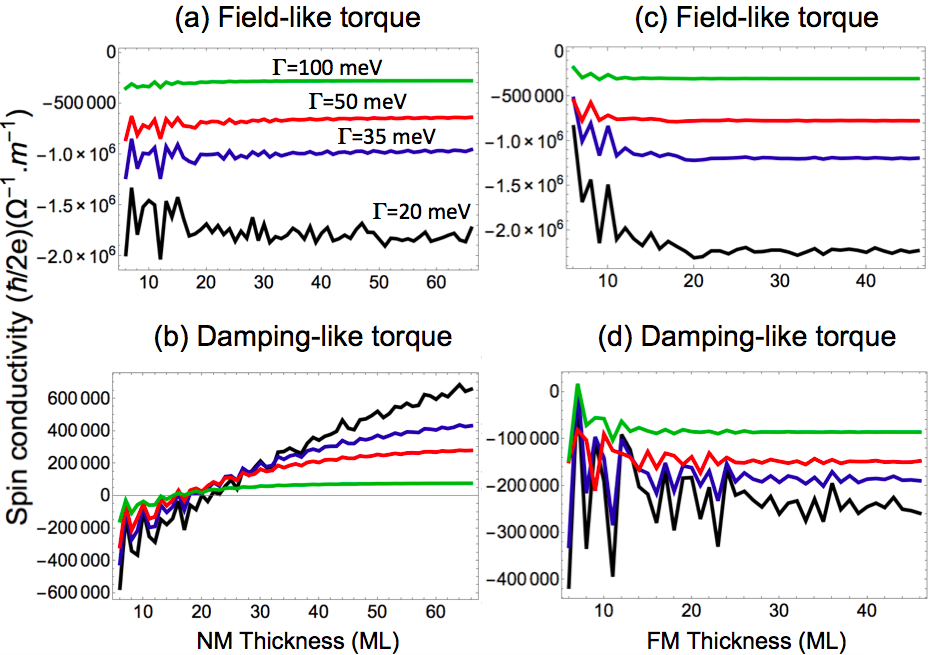}
      \caption{(Color online) Transport properties upon varying the nonmagnetic layer thickness (left panels) and the ferromagnetic layer thickness (right panels). This figure shows the thickness dependence of (a,c) the field-like torque and (b,d) the damping-like torque. The curves are calculated for a magnetization pointing along ${\bf z}$ and for various disorder strength, $\Gamma=10$ meV (black), $\Gamma=20$ meV (blue), $\Gamma=50$ meV (red), and $\Gamma=100$ meV (green).\label{Fig4}}
\end{figure}

\begin{figure}
        \includegraphics[width=6cm]{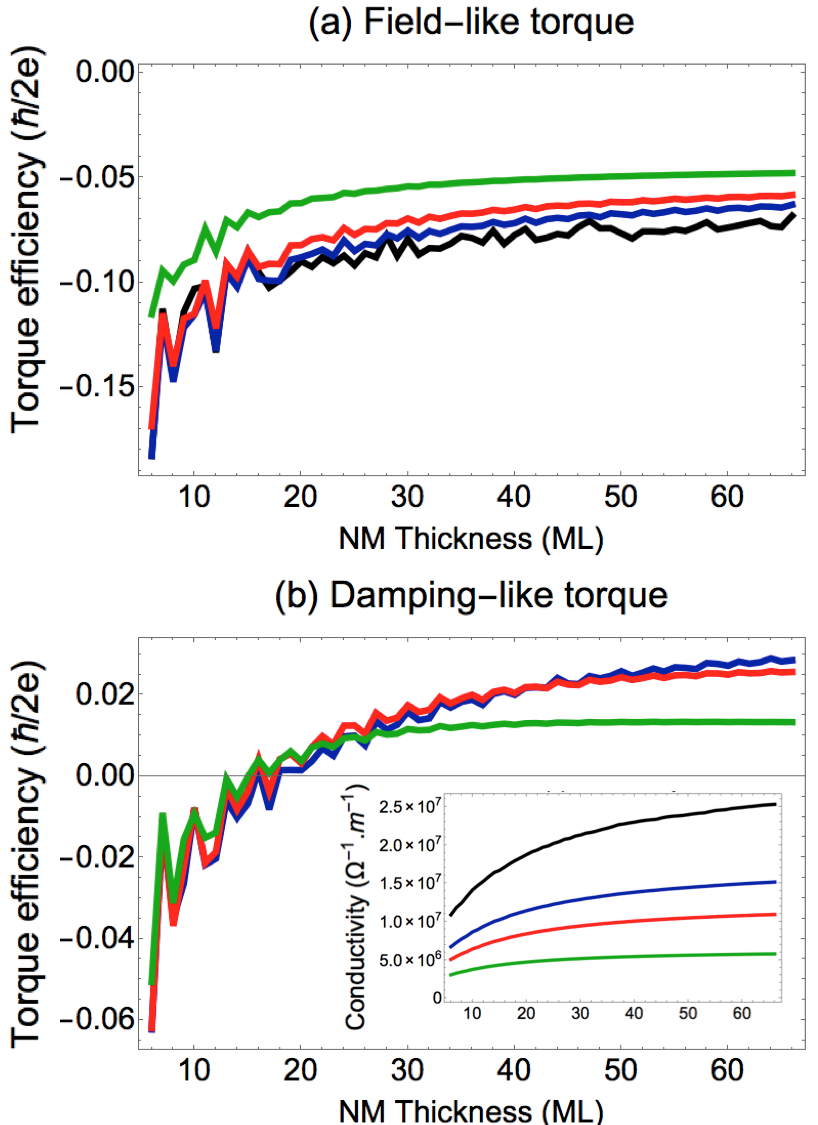}
      \caption{(Color online) Efficiency of the (a) field-like torque and (b) damping-like torque as a function of the nonmagnetic layer thickness. The curves are calculated for a magnetization pointing along ${\bf z}$ and for various disorder strength, $\Gamma=10$ meV (black), $\Gamma=20$ meV (blue), $\Gamma=50$ meV (red), and $\Gamma=100$ meV (green).\label{Fig4bis}}
\end{figure}

Figure \ref{Fig4} shows (a) field-like torque and (b) damping-like torque for various disorder strengths $\Gamma$ as a function of the thickness of the nonmagnetic metal. The corresponding conductivity is shown in the insert of Fig. \ref{Fig4bis}(b) for reference. It displays the usual $G_0/(1+3\lambda/8t)$ behavior expected in the semiclassical size effect regime \cite{Sondheimer2001}, which clearly indicates that the heterostructure doesn't enter the diffusive regime before the nonmagnetic layer thickness reaches about 10 nm, which is consistent with experimental reports \cite{Nguyen2016}. The field-like torque [Fig. \ref{Fig4}(a)] is mostly constant over the thickness range, displaying quantum oscillations over the first 20 monolayers ($\approx2.7$ nm) but keeping the same sign. In contrast, the damping-like torque [Fig. \ref{Fig4}(b)] progressively increases from a negative value to a positive one, before reaching saturation. The thickness at which the saturation is reached strongly depends on the disorder strength, suggesting that spin-dependent scattering plays an important role here. The change of sign occurs around 20 monolayers ($\approx2.7$ nm) and is weakly sensitive to the disorder, suggesting a transition between two "intrinsic" (i.e., band structure driven) mechanisms of opposite signs. This sign change is similar to the one observed experimentally \cite{Kim2013,Akyol2016,Ramaswamy2016}. Since our model does not account for complex scattering events, we suggest that this change of sign is associated with the competition between the interfacial Berry-curvature induced damping-like torque \cite{Kurebayashi2014} and the spin Hall effect coming from the bulk of the nonmagnetic material. Since the Berry-curvature induced damping-like torque is an interfacial effect, it does not significantly depend on the nonmagnetic metal thickness. On the contrary, the contribution to the damping-like torque from the spin Hall effect necessitates a nonmagnetic layer thickness larger than the spin relaxation length to be efficient and compensate the interfacial Berry-curvature induced contribution. One last remark is in order: in our simulation, the spin Hall and Berry-curvature induced contributions have opposite sign. However, we speculate this is only accidental as the spin Hall-driven contribution is controlled by the interplay between spin-orbit coupling and band filling as governed by Hund's third rule \cite{Tanaka2008,Freimuth2010}, whereas the interfacial Berry-curvature contribution is governed by the interfacial potential drop. This feature is therefore not general.\par

It is instructive to consider the thickness dependence of the torque efficiency, defined as the ratio between the torque and the conductivity of the heterostructure. This efficiency would be equivalent to the spin Hall angle in the case only spin Hall effect were present in the structure. The efficiency of the field-like and damping-like torques is reported on Fig. \ref{Fig4bis}(a) and (b), respectively, while the conductivity of the heterostructure is shown in the inset of (b), for reference. It is clear that the field-like torque efficiency is much larger for small thicknesses, as the current density is concentrated close to the interface. A similar feature is obtained for the damping-like torque efficiency. It is noticeable that the efficiency drops significantly within the first 10-15 monolayers ($\approx 2$ nm), showing that quantum confinement can be beneficial for spin-orbit torque. 

To complete this study, let us now consider the influence of the ferromagnetic layer thickness. Experimentally, it is found that the field-like component decreases strongly with the ferromagnetic layer thickness while the damping-like component remains mostly constant \cite{Kim2013}. We observe a similar feature in our calculations, shown in Fig. \ref{Fig4}(c) and (d). The field-like component increases upon increasing the ferromagnetic layer thickness and saturates after about 10 monolayers. The damping-like component displays a similar increase as a function of the ferromagnetic layer thickness, but it also exhibits large quantum oscillations, which makes the systematic increase more difficult to see at first glance. This behavior is associated with the absorption of the transverse spin current by the ferromagnetic layer over the spin dephasing length. If the ferromagnetic layer thickness is thinner than the spin dephasing length, the injected spin current (or, equivalently, the spin density smearing into the ferromagnetic layer) is not entirely absorbed and is reflected back into the nonmagnetic layer, resulting in a reduced torque. Upon increasing the ferromagnetic layer thickness, more spin current is absorbed, resulting in an increase and saturation of the torque (see, e.g., Ref. \onlinecite{Zwierzycki2005}). This scenario was experimentally confirmed recently \cite{Qiu2016}, but cannot be properly modeled using drift-diffusion theories due to the importance of quantum oscillations in this thickness range \cite{Haney2013b,Amin2016b}.

\subsection{Angular dependence}
The calculations presented above were all performed by setting the magnetization along ${\bf z}$. Yet, several experimental\cite{Garello2013,Qiu2015,Safranski2019} and theoretical studies\cite{Lee2015,Pauyac2013,Hals2014,Zelezny2017,Belashchenko2019} have pointed out that the spin-orbit torque does not reduce to the forms given in Eqs. \eqref{eq:fl}-\eqref{eq:dl}. For the highest $C_{\infty}$ symmetry, Belashchenko et al.\cite{Belashchenko2019} proposed that the spin-orbit torque be written
\begin{eqnarray}\label{eq:garello}
{\bf T}&=&-P_\theta^A{\bf m}\times({\bf z}\times{\bf E})+P_\theta^{A'}({\bf m}\cdot{\bf E}){\bf m}\times({\bf z}\times{\bf m})\\
&&-P_\theta^{B}{\bf m}\times[({\bf z}\times{\bf E})\times{\bf m}]+P_\theta^{B'}({\bf m}\cdot{\bf E}){\bf m}\times{\bf z}+...\nonumber
\end{eqnarray}
where $P_\theta^X=\sum_nX_{2n}P_{2n}(\cos\theta)$, $P_{2n}(x)$ being the Legendre polynomials. The first and third terms are simply the conventional field-like and damping-like torques. The second and fourth terms can be referred to as "planar" field-like and "planar" damping-like torques, respectively. These components are only non-zero when the magnetization lies along the applied electric field. In Ref. \onlinecite{Pauyac2013}, these two planar components were obtained analytically and related to the presence of D'yakonov-Perel' anisotropic spin relaxation. As a matter of fact, in such ultrathin magnetic heterostructures the spin component pointing perpendicular to the plane of the interface and that pointing in-plane relax at different rates, which modifies the overall spin dynamics at the interface, resulting in these additional torque components.\par

\begin{figure}
        \includegraphics[width=8.5cm]{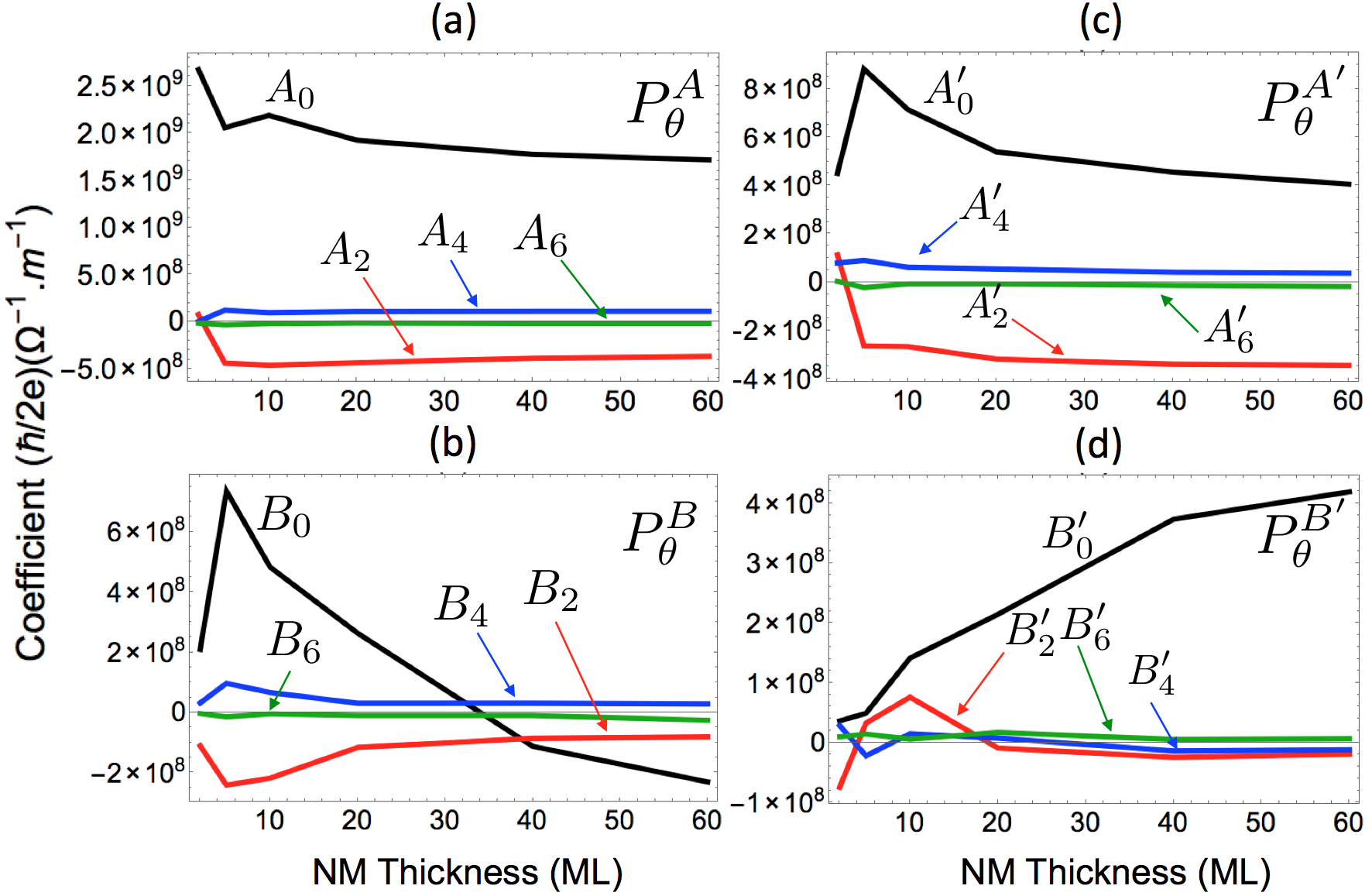}
      \caption{(Color online) Legendre expansion coefficients as function of the thickness of the nonmagnetic layer. These coefficients correspond to (a) the conventional field-like torque, (b) the conventional damping-like torque, (c) the planar field-like torque and (d) the planar damping-like torque.\label{fig:AngularFit}}
\end{figure}

To evaluate this angular anisotropy, we computed the two components $T_x$ and $T_y$ when varying the magnetization in the ($y,z$) and ($z,x$) planes. From Eq. \eqref{eq:garello}, we expect
\begin{eqnarray}\label{eq:garello2}
T_x/\cos\theta&=&P_\theta^A,\\
T_y/\cos^2\theta&=&P_\theta^B
\end{eqnarray}
when the magnetization rotates in the ($y,z$) plane, and 
\begin{eqnarray}\label{eq:garello3}
T_x/\cos\theta&=&P_\theta^A-\sin^2\theta P_\theta^{A'},\\
T_y&=&P_\theta^B+\sin^2\theta P_\theta^{B'}
\end{eqnarray}
where the magnetization rotates in the ($z,x$) plane. By fitting these angular dependences using Legendre polynomials, we obtain the first four components of the expansion of Eq. \eqref{eq:garello}. These components are reported on Fig. \ref{fig:AngularFit} upon varying the thickness of the nonmagnetic metal. 

The conventional field-like torque, reported in Fig. \ref{fig:AngularFit}(a), dominates all the other components and exhibits almost no angular dependence ($A_0\gg A_{2,4,6}$). The conventional damping-like torque, reported in Fig. \ref{fig:AngularFit}(b), is about one order of magnitude smaller, exhibits the sign reversal discussed previously and displays a sizable angular dependence at small thicknesses ($B_2\gg B_{4,6}$). This angular dependence vanishes upon increasing the thickness of the nonmagnetic layer. Interestingly, the two "planar" components exhibit a radically different behavior. First of all, both components are comparable in magnitude with the damping-like torque, which means that they play a crucial role in current-driven dynamics and cannot be neglected. Second, the planar field-like torque [Fig. \ref{fig:AngularFit}(c)] exhibits a substantial angular dependence ($A_{0}'\approx A_{2}'\gg A'_{4,6}$) that saturates after a few monolayers only. This indicates that this component is mostly of interfacial origin, in agreement with the D'yakonov-Perel' scenario evoked in Ref. \onlinecite{Pauyac2013}. Finally, the planar damping-like torque presents a surprising behavior [Fig. \ref{fig:AngularFit}(d)]. It displays almost no angular dependence (except at small thicknesses), and increases steadily over a few tens of monolayers before reaching saturation at large thicknesses. This progressive saturation is similar to the one expected for spin Hall-driven damping torque originating from the nonmagnetic layer, as mentioned in Section \ref{s:thickn}. 

Therefore, the results reported on Fig. \ref{fig:AngularFit} suggest that the planar field-like torque is associated with an interfacial effect and can be seen at the companion of the conventional interfacial (Rashba) field-like torque, while the planar damping-like torque is associated with bulk mechanisms and accompanies the conventional (spin Hall-driven) damping-like torque. To complete this discussion we emphasize that \citet{Safranski2019} reported a planar Hall torque that they attributed to the planar Hall effect from the bulk of the ferromagnet. In our case, the spin-orbit coupling of the ferromagnet remains quite small (see Section \ref{s:self}) and it is unlikely that such a mechanism contributes to the torques reported on Fig. \ref{fig:AngularFit}.

\section{Self-torque in the ferromagnet \label{s:self}}

\begin{figure}
        \includegraphics[width=8cm]{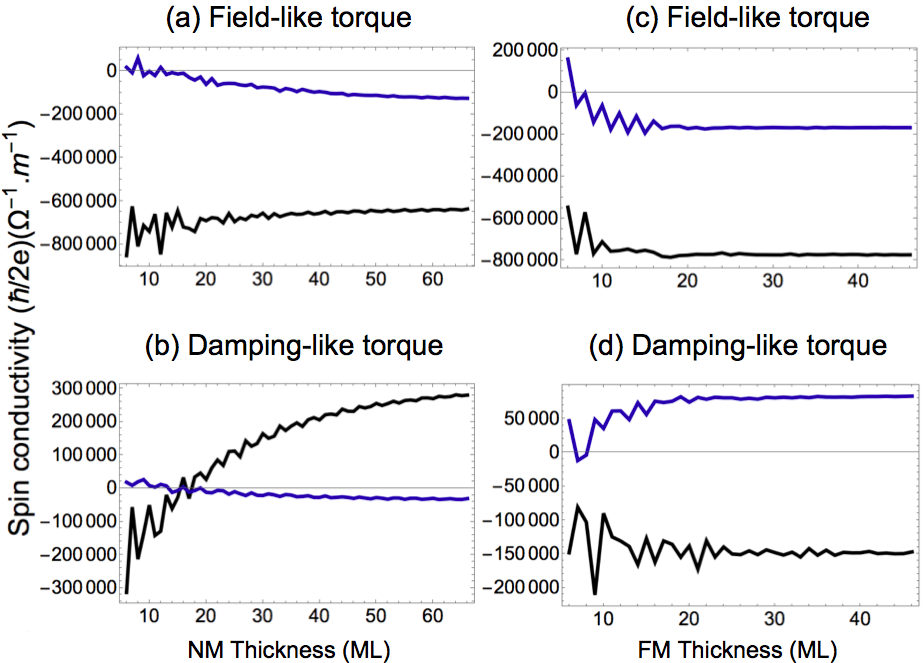}
      \caption{(Color online) Spin-orbit torque components upon varying the nonmagnetic layer thickness (left panels) and the ferromagnetic layer thickness (right panels), when spin-orbit coupling is present in both nonmagnetic and ferromagnetic layers (black) and when it is present only in the ferromagnetic layer (blue). The curves are calculated for a magnetization pointing along ${\bf z}$ and for $\Gamma=50$ meV.\label{Fig10}}
\end{figure}

To complete this study, we now turn off the spin-orbit coupling of the nonmagnetic layer. The spatial profile of the spin density is shown in Fig. \ref{Fig2}, red curves. The features described above survive: $S_x$ is more delocalized than $S_y$, although their magnitude is much (three or four times) weaker than in the case where spin-orbit coupling is present in both layers. The thickness dependence is shown in Fig. \ref{Fig10}, blue curves. The black curves represent the case where the spin-orbit coupling is present in both layers and serves as a reference. The field-like torque starts slightly positive [Fig. \ref{Fig10}(a)], switches sign around about 10 monolayers and increases negatively until reaching saturation at about 50 monolayers. The damping-like torque shows a similar behavior. It also starts slightly positive [Fig. \ref{Fig10}(b)], switches sign about 15 monolayers and increases negatively until reaching saturation at about 60 monolayers. It is interesting to note that the self-field-like torque has the same sign as the case where spin-orbit coupling is present everywhere, whereas the self-damping-like torque is opposite. 

\begin{figure}
        \includegraphics[width=9cm]{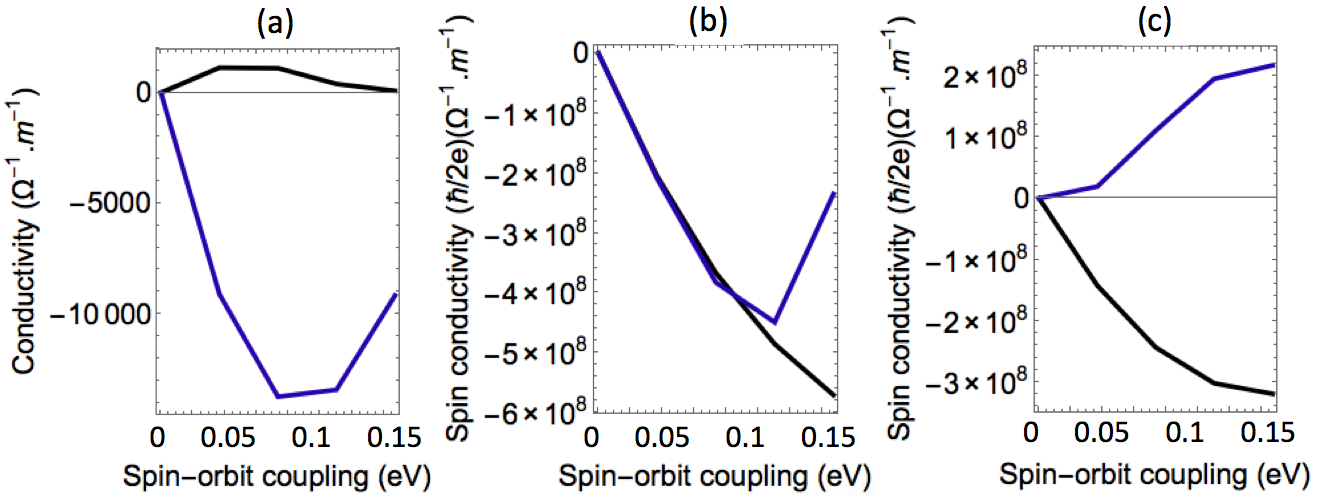}
      \caption{(Color online) Dependence of the (a) anomalous Hall conductivity, (b) field-like and (c) damping-like components of the spin-orbit torque upon varying the spin-orbit coupling $\xi_{\rm so}$. The black lines represent the case where the spin-orbit coupling of the ferromagnet is set to zero, whereas the blue lines represent the case where the spin-orbit coupling of the nonmagnetic metal is set to zero. In this calculation, we set the nonmagnetic metal thickness to 40 monolayers and the ferromagnet thickness to 7 monolayers. The curves are calculated for a magnetization pointing along ${\bf z}$ and for $\Gamma=50$ meV.\label{Fig11}}
\end{figure}

To understand this distinct behavior, we compute the dependence of the anomalous Hall conductivity and torque components as a function of the spin-orbit coupling energy of the individual layers. The results are reported on Fig. \ref{Fig11}. The black lines correspond to the case where the spin-orbit coupling is in the nonmagnetic metal only, whereas the blue lines correspond to the case where the spin-orbit coupling is in the ferromagnet only. The anomalous Hall conductivity [Fig. \ref{Fig11}(a)] and damping-like torque [Fig. \ref{Fig11}(c)] both change sign depending on which layer possesses spin-orbit coupling. In contrast, the field-like torque remains negative, irrespective of where the spin-orbit coupling is [Fig. \ref{Fig11}(b)].\par

The field-like torque, as explained in Section \ref{s:porbitals}, is associated with the interfacial, Rashba-like spin-orbit coupling, whose sign is governed by the interfacial potential drop. Therefore, for a given spin-orbit coupling strength, its sign is opposite on the two sides of the interface [see Eq. \eqref{eq:rashba}]. This seems contradictory with the results of Fig. \ref{Fig11}(b) and suggests that the sign of the spin-orbit coupling experienced by the Bloch states of the nonmagnetic layer is opposite to the one experienced by the Bloch states of the ferromagnet. This observation is consistent with Hund's third rule that states that for materials with more-than-half-filled electronic shells such as Fe, the spin and orbital momenta are aligned with each other, while for materials with less-than-half-filled electronic shells like W, there are anti-aligned. Since our tight-binding model is parameterized on these two elements, it is reasonable that Hund's third rule applies. As a consequence, the opposite potential drop felt by Bloch states on each side of the interface is compensated by the opposite effective spin-orbit coupling, and the field-like torque is the same whether the spin-orbit coupling is on the ferromagnet or on the nonmagnetic layer.\par

In contrast, the damping-like torque at large thicknesses and the anomalous Hall conductivity are not associated with interfacial potential drop, but rather with the (spin) Berry curvature of the bulk material. It is therefore solely governed by the effective spin-orbit coupling experienced by the Bloch electrons and is opposite when switching the spin-orbit coupling from the ferromagnet to the nonmagnetic metal.

The dependence as a function of the ferromagnetic layer thickness is reported in Fig. \ref{Fig10}(c) and (d) for the field-like and damping-like torques, respectively. We obtain similar thickness dependence as in the case where spin-orbit coupling is present in both layers, reflecting the importance of the spin dephasing length. Using our realistic parameters, the self-torque we obtain is about four to fives times smaller in magnitude compared to the torque arising from the nonmagnetic metal, consistent with the relative magnitude of the spin-orbit coupling (about 65 meV in Fe compared to 360 meV in W). These calculations support an idea that was put forward in Ref. \onlinecite{Pauyac2018}: the spin Hall current generated inside the ferromagnetic layer can create an efficient torque on the magnetic order as long as the two opposite interfaces are dissimilar.

\section{Conclusion}
Using a multi-orbital tight-binding model, we computed the spin-orbit torque in a transition metal heterostructure, treating bulk and interfacial spin-orbit effects coherently and on equal footing. Thickness and angular dependences of the torque show that it possesses four sizable components, the conventional field-like and damping-like torques, as well as two planar components that vanish when the magnetization lies out-of-plane. The conventional field-like torque is entirely controlled by the interface, as expected from interfacial inverse spin galvanic effect, while the damping-like torque possesses two components, an interfacial one dominating at small thicknesses and a bulk contribution dominating at large thicknesses. The former is attributed to the intrinsic interfacial Berry-curvature-driven damping torque \cite{Kurebayashi2014}, whereas the latter is associated with the spin Hall effect generated in the bulk of the nonmagnetic metal.\par

Interestingly, the planar field-like torque shows substantial angular dependence and is of interfacial origin, like the conventional field-like torque. In contrast, the planar damping-like torque does not exhibit angular dependence and increases with the nonmagnetic metal thickness, indicating that it originates from the bulk of the nonmagnetic layer, similarly to the conventional spin Hall-driven damping torque. Our results demonstrate that these four torque components are present in any transition metal heterostructures and must be taken into account when interpreting the experimental data, and in particular the current-driven magnetization dynamics.\par

Finally, we investigate the self-torque exerted on the ferromagnet when spin-orbit coupling of the nonmagnetic metal is turned off. Our results suggest that the spin accumulation that builds up inside the ferromagnet can be large enough to induce magnetization excitations.

\acknowledgments
This work was supported by the King Abdullah University of Science and Technology (KAUST) through the Office of
Sponsored Research (OSR) [Grant Number OSR-2017-CRG6-3390].  

\appendix*
\section{Hopping integrals}
The parameters appearing in Eq. \eqref{eq:Hmono} are
\begin{eqnarray*}
&&\gamma_{xy}^{\bf k}=\varepsilon_{xy}+(3V^2_\sigma +V^2_\pi)\cos\tilde{k}_x\cos\tilde{k}_y,\\
&&\gamma_{yz}^{\bf k}=\varepsilon_{yz}+2(V^2_\pi +V^2_\delta)\cos\tilde{k}_x\cos\tilde{k}_y,\\
&&\gamma_{zx}^{\bf k}=\varepsilon_{zx}+2(V^2_\pi +V^2_\delta)\cos\tilde{k}_x\cos\tilde{k}_y,\\
&&\gamma_{z^2}^{\bf k}=\varepsilon_{z^2}+(V^2_\sigma +3V^2_\delta)\cos\tilde{k}_x\cos\tilde{k}_y,\\
&&\gamma_{x^2-y^2}^{\bf k}=\varepsilon_{yz}+4V^2_\pi\cos\tilde{k}_x\cos\tilde{k}_y,\\
&&t_{zx,yz}^{\bf k} =2(V^2_\delta-V^2_\pi)\sin\tilde{k}_x\sin\tilde{k}_y,\\
&&t_{xy,z^2}^{\bf k} =\sqrt{3}(V^2_\sigma-V^2_\pi)\sin\tilde{k}_x\sin\tilde{k}_y,
\end{eqnarray*}
where $V^2_{\sigma,\pi,\delta}$ are the two-site second nearest neighbor hopping integrals in the ($x,y$) plane \cite{Slater1954}, as depicted in Fig. \ref{Fig0}(b), and $\tilde{k}_{x,y}=k_{x,y}a_0/2$. The matrix elements of the nearest neighbor hopping matrix ${\cal T}_1$ are 
\begin{eqnarray*}
&&t_{xy}^{z,{\bf k}}=(V^1_\pi +V^1_\delta)\left(\cos\tilde{k}_x+\cos\tilde{k}_y\right),\\
&&t_{yz}^{z,{\bf k}}=(V^1_\pi +V^1_\delta)\cos\tilde{k}_x+2(V^1_\sigma+V^1_\delta)\cos\tilde{k}_y,\\
&&t_{zx}^{z,{\bf k}}=(V^1_\pi +V^1_\delta)\cos\tilde{k}_y+2(V^1_\sigma+V^1_\delta)\cos\tilde{k}_x,\\
&&t_{x^2-y^2}^{z,{\bf k}}=\frac{1}{8}(3V^1_\sigma+4V^1_\pi +9V^1_\delta)\left(\cos\tilde{k}_x+\cos\tilde{k}_y\right),\\
&&t_{z^2}^{z,{\bf k}}=\frac{1}{8}(V^1_\sigma+12V^1_\pi +3V^1_\delta)\left(\cos\tilde{k}_x+\cos\tilde{k}_y\right),
\end{eqnarray*}
where we used the shorthand notation $t_{\nu,\nu}^{z,{\bf k}}=t_{\nu}^{z,{\bf k}}$, and
\begin{eqnarray*}
&&t_{xy,zx}^{z,{\bf k}}=-i(V^1_\pi -V^1_\delta)\sin\tilde{k}_y,\\
&&t_{xy,yz}^{z,{\bf k}}=-i(V^1_\pi -V^1_\delta)\sin\tilde{k}_x,\\
&&t_{yz,x^2-y^2}^{z,{\bf k}}=\frac{3i}{4}(V^1_\sigma -V^1_\delta)\sin\tilde{k}_y,\\
&&t_{zx,x^2-y^2}^{z,{\bf k}}=-\frac{3i}{4}(V^1_\sigma -V^1_\delta)\sin\tilde{k}_x,\\
&&t_{z^2,x^2-y^2}^{z,{\bf k}}=\frac{\sqrt{3}}{8}(V^1_\sigma +3V^1_\delta-4V^1_\pi)\left(\cos\tilde{k}_x-\cos\tilde{k}_y\right),\\
&&t_{yz,z^2}^{z,{\bf k}}=-\frac{i\sqrt{3}}{4}(V^1_\sigma -V^1_\delta)\sin\tilde{k}_y,\\
&&t_{zx,z^2}^{z,{\bf k}}=-\frac{i\sqrt{3}}{4}(V^1_\sigma -V^1_\delta)\sin\tilde{k}_x,\\
&&t_{yz,zx}^{z,{\bf k}}=t_{xy,x^2-y^2}^{z,{\bf k}}=t_{xy,z^2}^{z,{\bf k}}=0.
\end{eqnarray*}
Again, $V^1_{\sigma,\pi,\delta}$ are the two-site nearest neighbor hopping integrals\cite{Slater1954}, as depicted in Fig. \ref{Fig0}(b). The second nearest neighbor hopping matrix ${\cal T}_2$ reads
\begin{eqnarray}\label{eq:t2}
{\cal T}_2={\rm Diag}\left(V^2_\delta,V^2_\pi,V^2_\pi,V^2_\sigma,V^2_\delta\right),
\end{eqnarray}
which connects the $n$-th monolayer to the $n+2$-th monolayer. 

\bibliography{Biblio2019}

\begin{thebibliography}{94}%
\makeatletter
\providecommand \@ifxundefined [1]{%
 \@ifx{#1\undefined}
}%
\providecommand \@ifnum [1]{%
 \ifnum #1\expandafter \@firstoftwo
 \else \expandafter \@secondoftwo
 \fi
}%
\providecommand \@ifx [1]{%
 \ifx #1\expandafter \@firstoftwo
 \else \expandafter \@secondoftwo
 \fi
}%
\providecommand \natexlab [1]{#1}%
\providecommand \enquote  [1]{``#1''}%
\providecommand \bibnamefont  [1]{#1}%
\providecommand \bibfnamefont [1]{#1}%
\providecommand \citenamefont [1]{#1}%
\providecommand \href@noop [0]{\@secondoftwo}%
\providecommand \href [0]{\begingroup \@sanitize@url \@href}%
\providecommand \@href[1]{\@@startlink{#1}\@@href}%
\providecommand \@@href[1]{\endgroup#1\@@endlink}%
\providecommand \@sanitize@url [0]{\catcode `\\12\catcode `\$12\catcode
  `\&12\catcode `\#12\catcode `\^12\catcode `\_12\catcode `\%12\relax}%
\providecommand \@@startlink[1]{}%
\providecommand \@@endlink[0]{}%
\providecommand \url  [0]{\begingroup\@sanitize@url \@url }%
\providecommand \@url [1]{\endgroup\@href {#1}{\urlprefix }}%
\providecommand \urlprefix  [0]{URL }%
\providecommand \Eprint [0]{\href }%
\providecommand \doibase [0]{http://dx.doi.org/}%
\providecommand \selectlanguage [0]{\@gobble}%
\providecommand \bibinfo  [0]{\@secondoftwo}%
\providecommand \bibfield  [0]{\@secondoftwo}%
\providecommand \translation [1]{[#1]}%
\providecommand \BibitemOpen [0]{}%
\providecommand \bibitemStop [0]{}%
\providecommand \bibitemNoStop [0]{.\EOS\space}%
\providecommand \EOS [0]{\spacefactor3000\relax}%
\providecommand \BibitemShut  [1]{\csname bibitem#1\endcsname}%
\let\auto@bib@innerbib\@empty
\bibitem [{\citenamefont {Brataas}\ and\ \citenamefont
  {Hals}(2014)}]{Brataas2014}%
  \BibitemOpen
  \bibfield  {author} {\bibinfo {author} {\bibfnamefont {A.}~\bibnamefont
  {Brataas}}\ and\ \bibinfo {author} {\bibfnamefont {K.~M.~D.}\ \bibnamefont
  {Hals}},\ }\href {\doibase 10.1038/nnano.2014.8} {\bibfield  {journal}
  {\bibinfo  {journal} {Nature Nanotechnology}\ }\textbf {\bibinfo {volume}
  {9}},\ \bibinfo {pages} {86} (\bibinfo {year} {2014})}\BibitemShut {NoStop}%
\bibitem [{\citenamefont {Manchon}\ \emph {et~al.}(2019)\citenamefont
  {Manchon}, \citenamefont {Zelezn{\'{y}}}, \citenamefont {Miron},
  \citenamefont {Jungwirth}, \citenamefont {Sinova}, \citenamefont {Thiaville},
  \citenamefont {Garello},\ and\ \citenamefont {Gambardella}}]{Manchon2019}%
  \BibitemOpen
  \bibfield  {author} {\bibinfo {author} {\bibfnamefont {A.}~\bibnamefont
  {Manchon}}, \bibinfo {author} {\bibfnamefont {J.}~\bibnamefont
  {Zelezn{\'{y}}}}, \bibinfo {author} {\bibfnamefont {M.}~\bibnamefont
  {Miron}}, \bibinfo {author} {\bibfnamefont {T.}~\bibnamefont {Jungwirth}},
  \bibinfo {author} {\bibfnamefont {J.}~\bibnamefont {Sinova}}, \bibinfo
  {author} {\bibfnamefont {A.}~\bibnamefont {Thiaville}}, \bibinfo {author}
  {\bibfnamefont {K.}~\bibnamefont {Garello}}, \ and\ \bibinfo {author}
  {\bibfnamefont {P.}~\bibnamefont {Gambardella}},\ }\href {\doibase
  10.1103/RevModPhys.91.035004} {\bibfield  {journal} {\bibinfo  {journal}
  {Review of Modern Physics}\ }\textbf {\bibinfo {volume} {91}},\ \bibinfo
  {pages} {035004} (\bibinfo {year} {2019})}\BibitemShut {NoStop}%
\bibitem [{\citenamefont {Miron}\ \emph {et~al.}(2011)\citenamefont {Miron},
  \citenamefont {Garello}, \citenamefont {Gaudin}, \citenamefont {Zermatten},
  \citenamefont {Costache}, \citenamefont {Auffret}, \citenamefont {Bandiera},
  \citenamefont {Rodmacq}, \citenamefont {Schuhl},\ and\ \citenamefont
  {Gambardella}}]{Miron2011b}%
  \BibitemOpen
  \bibfield  {author} {\bibinfo {author} {\bibfnamefont {I.~M.}\ \bibnamefont
  {Miron}}, \bibinfo {author} {\bibfnamefont {K.}~\bibnamefont {Garello}},
  \bibinfo {author} {\bibfnamefont {G.}~\bibnamefont {Gaudin}}, \bibinfo
  {author} {\bibfnamefont {P.~J.}\ \bibnamefont {Zermatten}}, \bibinfo {author}
  {\bibfnamefont {M.~V.}\ \bibnamefont {Costache}}, \bibinfo {author}
  {\bibfnamefont {S.}~\bibnamefont {Auffret}}, \bibinfo {author} {\bibfnamefont
  {S.}~\bibnamefont {Bandiera}}, \bibinfo {author} {\bibfnamefont
  {B.}~\bibnamefont {Rodmacq}}, \bibinfo {author} {\bibfnamefont
  {A.}~\bibnamefont {Schuhl}}, \ and\ \bibinfo {author} {\bibfnamefont
  {P.}~\bibnamefont {Gambardella}},\ }\href {\doibase 10.1038/nature10309}
  {\bibfield  {journal} {\bibinfo  {journal} {Nature}\ }\textbf {\bibinfo
  {volume} {476}},\ \bibinfo {pages} {189} (\bibinfo {year}
  {2011})}\BibitemShut {NoStop}%
\bibitem [{\citenamefont {Liu}\ \emph {et~al.}(2012)\citenamefont {Liu},
  \citenamefont {Pai}, \citenamefont {Li}, \citenamefont {Tseng}, \citenamefont
  {Ralph},\ and\ \citenamefont {Buhrman}}]{Liu2012}%
  \BibitemOpen
  \bibfield  {author} {\bibinfo {author} {\bibfnamefont {L.}~\bibnamefont
  {Liu}}, \bibinfo {author} {\bibfnamefont {C.-F.}\ \bibnamefont {Pai}},
  \bibinfo {author} {\bibfnamefont {Y.}~\bibnamefont {Li}}, \bibinfo {author}
  {\bibfnamefont {H.~W.}\ \bibnamefont {Tseng}}, \bibinfo {author}
  {\bibfnamefont {D.~C.}\ \bibnamefont {Ralph}}, \ and\ \bibinfo {author}
  {\bibfnamefont {R.~A.}\ \bibnamefont {Buhrman}},\ }\href {\doibase
  10.1126/science.1218197} {\bibfield  {journal} {\bibinfo  {journal}
  {Science}\ }\textbf {\bibinfo {volume} {336}},\ \bibinfo {pages} {555}
  (\bibinfo {year} {2012})}\BibitemShut {NoStop}%
\bibitem [{\citenamefont {Sinova}\ \emph {et~al.}(2015)\citenamefont {Sinova},
  \citenamefont {Valenzuela}, \citenamefont {Wunderlich}, \citenamefont
  {Back},\ and\ \citenamefont {Jungwirth}}]{Sinova2015}%
  \BibitemOpen
  \bibfield  {author} {\bibinfo {author} {\bibfnamefont {J.}~\bibnamefont
  {Sinova}}, \bibinfo {author} {\bibfnamefont {S.~O.}\ \bibnamefont
  {Valenzuela}}, \bibinfo {author} {\bibfnamefont {J.}~\bibnamefont
  {Wunderlich}}, \bibinfo {author} {\bibfnamefont {C.~H.}\ \bibnamefont
  {Back}}, \ and\ \bibinfo {author} {\bibfnamefont {T.}~\bibnamefont
  {Jungwirth}},\ }\href@noop {} {\bibfield  {journal} {\bibinfo  {journal}
  {Review of Modern Physics}\ }\textbf {\bibinfo {volume} {87}},\ \bibinfo
  {pages} {1213} (\bibinfo {year} {2015})}\BibitemShut {NoStop}%
\bibitem [{\citenamefont {Manchon}\ and\ \citenamefont
  {Zhang}(2008)}]{Manchon2008b}%
  \BibitemOpen
  \bibfield  {author} {\bibinfo {author} {\bibfnamefont {A.}~\bibnamefont
  {Manchon}}\ and\ \bibinfo {author} {\bibfnamefont {S.}~\bibnamefont
  {Zhang}},\ }\href@noop {} {\bibfield  {journal} {\bibinfo  {journal}
  {Physical Review B}\ }\textbf {\bibinfo {volume} {78}},\ \bibinfo {pages}
  {212405} (\bibinfo {year} {2008})}\BibitemShut {NoStop}%
\bibitem [{\citenamefont {Manchon}\ and\ \citenamefont
  {Zhang}(2009)}]{Manchon2009b}%
  \BibitemOpen
  \bibfield  {author} {\bibinfo {author} {\bibfnamefont {A.}~\bibnamefont
  {Manchon}}\ and\ \bibinfo {author} {\bibfnamefont {S.}~\bibnamefont
  {Zhang}},\ }\href {\doibase 10.1103/PhysRevB.79.094422} {\bibfield  {journal}
  {\bibinfo  {journal} {Physical Review B}\ }\textbf {\bibinfo {volume} {79}},\
  \bibinfo {pages} {094422} (\bibinfo {year} {2009})}\BibitemShut {NoStop}%
\bibitem [{\citenamefont {Garate}\ and\ \citenamefont
  {MacDonald}(2009)}]{Garate2009}%
  \BibitemOpen
  \bibfield  {author} {\bibinfo {author} {\bibfnamefont {I.}~\bibnamefont
  {Garate}}\ and\ \bibinfo {author} {\bibfnamefont {A.~H.}\ \bibnamefont
  {MacDonald}},\ }\href {\doibase 10.1103/PhysRevB.80.134403} {\bibfield
  {journal} {\bibinfo  {journal} {Physical Review B}\ }\textbf {\bibinfo
  {volume} {80}},\ \bibinfo {pages} {134403} (\bibinfo {year}
  {2009})}\BibitemShut {NoStop}%
\bibitem [{\citenamefont {Lifshits}\ and\ \citenamefont
  {Dyakonov}(2009)}]{Lifshits2009}%
  \BibitemOpen
  \bibfield  {author} {\bibinfo {author} {\bibfnamefont {M.~B.}\ \bibnamefont
  {Lifshits}}\ and\ \bibinfo {author} {\bibfnamefont {M.~I.}\ \bibnamefont
  {Dyakonov}},\ }\href {\doibase 10.1103/PhysRevLett.103.186601} {\bibfield
  {journal} {\bibinfo  {journal} {Physical Review Letters}\ }\textbf {\bibinfo
  {volume} {103}},\ \bibinfo {pages} {186601} (\bibinfo {year}
  {2009})}\BibitemShut {NoStop}%
\bibitem [{\citenamefont {Saidaoui}\ \emph {et~al.}(2015)\citenamefont
  {Saidaoui}, \citenamefont {Otani},\ and\ \citenamefont
  {Manchon}}]{Saidaoui2015b}%
  \BibitemOpen
  \bibfield  {author} {\bibinfo {author} {\bibfnamefont {H.}~\bibnamefont
  {Saidaoui}}, \bibinfo {author} {\bibfnamefont {Y.}~\bibnamefont {Otani}}, \
  and\ \bibinfo {author} {\bibfnamefont {A.}~\bibnamefont {Manchon}},\ }\href
  {\doibase 10.1103/PhysRevB.92.024417} {\bibfield  {journal} {\bibinfo
  {journal} {Physical Review B}\ }\textbf {\bibinfo {volume} {92}},\ \bibinfo
  {pages} {024417} (\bibinfo {year} {2015})}\BibitemShut {NoStop}%
\bibitem [{\citenamefont {Saidaoui}\ and\ \citenamefont
  {Manchon}(2016)}]{Saidaoui2016}%
  \BibitemOpen
  \bibfield  {author} {\bibinfo {author} {\bibfnamefont {H.}~\bibnamefont
  {Saidaoui}}\ and\ \bibinfo {author} {\bibfnamefont {A.}~\bibnamefont
  {Manchon}},\ }\href {\doibase 10.1103/PhysRevLett.117.036601} {\bibfield
  {journal} {\bibinfo  {journal} {Physical Review Letters}\ }\textbf {\bibinfo
  {volume} {117}},\ \bibinfo {pages} {036601} (\bibinfo {year}
  {2016})}\BibitemShut {NoStop}%
\bibitem [{\citenamefont {Amin}\ \emph {et~al.}(2018)\citenamefont {Amin},
  \citenamefont {Zemen},\ and\ \citenamefont {Stiles}}]{Amin2018}%
  \BibitemOpen
  \bibfield  {author} {\bibinfo {author} {\bibfnamefont {V.~P.}\ \bibnamefont
  {Amin}}, \bibinfo {author} {\bibfnamefont {J.}~\bibnamefont {Zemen}}, \ and\
  \bibinfo {author} {\bibfnamefont {M.~D.}\ \bibnamefont {Stiles}},\ }\href
  {\doibase 10.1103/PhysRevLett.121.136805} {\bibfield  {journal} {\bibinfo
  {journal} {Physical Review Letters}\ }\textbf {\bibinfo {volume} {121}},\
  \bibinfo {pages} {136805} (\bibinfo {year} {2018})}\BibitemShut {NoStop}%
\bibitem [{\citenamefont {Freimuth}\ \emph {et~al.}(2018)\citenamefont
  {Freimuth}, \citenamefont {Bl{\"{u}}gel},\ and\ \citenamefont
  {Mokrousov}}]{Freimuth2018}%
  \BibitemOpen
  \bibfield  {author} {\bibinfo {author} {\bibfnamefont {F.}~\bibnamefont
  {Freimuth}}, \bibinfo {author} {\bibfnamefont {S.}~\bibnamefont
  {Bl{\"{u}}gel}}, \ and\ \bibinfo {author} {\bibfnamefont {Y.}~\bibnamefont
  {Mokrousov}},\ }\href {\doibase 10.1103/PhysRevB.98.024419} {\bibfield
  {journal} {\bibinfo  {journal} {Physical Review B}\ }\textbf {\bibinfo
  {volume} {98}},\ \bibinfo {pages} {024419} (\bibinfo {year}
  {2018})}\BibitemShut {NoStop}%
\bibitem [{\citenamefont {Kim}\ \emph {et~al.}(2013)\citenamefont {Kim},
  \citenamefont {Sinha}, \citenamefont {Hayashi}, \citenamefont {Yamanouchi},
  \citenamefont {Fukami}, \citenamefont {Suzuki}, \citenamefont {Mitani},\ and\
  \citenamefont {Ohno}}]{Kim2013}%
  \BibitemOpen
  \bibfield  {author} {\bibinfo {author} {\bibfnamefont {J.}~\bibnamefont
  {Kim}}, \bibinfo {author} {\bibfnamefont {J.}~\bibnamefont {Sinha}}, \bibinfo
  {author} {\bibfnamefont {M.}~\bibnamefont {Hayashi}}, \bibinfo {author}
  {\bibfnamefont {M.}~\bibnamefont {Yamanouchi}}, \bibinfo {author}
  {\bibfnamefont {S.}~\bibnamefont {Fukami}}, \bibinfo {author} {\bibfnamefont
  {T.}~\bibnamefont {Suzuki}}, \bibinfo {author} {\bibfnamefont
  {S.}~\bibnamefont {Mitani}}, \ and\ \bibinfo {author} {\bibfnamefont
  {H.}~\bibnamefont {Ohno}},\ }\href {\doibase 10.1038/nmat3522} {\bibfield
  {journal} {\bibinfo  {journal} {Nature Materials}\ }\textbf {\bibinfo
  {volume} {12}},\ \bibinfo {pages} {240} (\bibinfo {year} {2013})}\BibitemShut
  {NoStop}%
\bibitem [{\citenamefont {Garello}\ \emph {et~al.}(2013)\citenamefont
  {Garello}, \citenamefont {Miron}, \citenamefont {Avci}, \citenamefont
  {Freimuth}, \citenamefont {Mokrousov}, \citenamefont {Bl{\"{u}}gel},
  \citenamefont {Auffret}, \citenamefont {Boulle}, \citenamefont {Gaudin},\
  and\ \citenamefont {Gambardella}}]{Garello2013}%
  \BibitemOpen
  \bibfield  {author} {\bibinfo {author} {\bibfnamefont {K.}~\bibnamefont
  {Garello}}, \bibinfo {author} {\bibfnamefont {I.~M.}\ \bibnamefont {Miron}},
  \bibinfo {author} {\bibfnamefont {C.~O.}\ \bibnamefont {Avci}}, \bibinfo
  {author} {\bibfnamefont {F.}~\bibnamefont {Freimuth}}, \bibinfo {author}
  {\bibfnamefont {Y.}~\bibnamefont {Mokrousov}}, \bibinfo {author}
  {\bibfnamefont {S.}~\bibnamefont {Bl{\"{u}}gel}}, \bibinfo {author}
  {\bibfnamefont {S.}~\bibnamefont {Auffret}}, \bibinfo {author} {\bibfnamefont
  {O.}~\bibnamefont {Boulle}}, \bibinfo {author} {\bibfnamefont
  {G.}~\bibnamefont {Gaudin}}, \ and\ \bibinfo {author} {\bibfnamefont
  {P.}~\bibnamefont {Gambardella}},\ }\href {\doibase 10.1038/nnano.2013.145}
  {\bibfield  {journal} {\bibinfo  {journal} {Nature Nanotechnology}\ }\textbf
  {\bibinfo {volume} {8}},\ \bibinfo {pages} {587} (\bibinfo {year}
  {2013})}\BibitemShut {NoStop}%
\bibitem [{\citenamefont {Avci}\ \emph {et~al.}(2014)\citenamefont {Avci},
  \citenamefont {Garello}, \citenamefont {Nistor}, \citenamefont {Godey},
  \citenamefont {Ballesteros}, \citenamefont {Mugarza}, \citenamefont {Barla},
  \citenamefont {Valvidares}, \citenamefont {Pellegrin}, \citenamefont {Ghosh},
  \citenamefont {Miron}, \citenamefont {Boulle}, \citenamefont {Auffret},
  \citenamefont {Gaudin},\ and\ \citenamefont {Gambardella}}]{Avci2014}%
  \BibitemOpen
  \bibfield  {author} {\bibinfo {author} {\bibfnamefont {C.~O.}\ \bibnamefont
  {Avci}}, \bibinfo {author} {\bibfnamefont {K.}~\bibnamefont {Garello}},
  \bibinfo {author} {\bibfnamefont {C.}~\bibnamefont {Nistor}}, \bibinfo
  {author} {\bibfnamefont {S.}~\bibnamefont {Godey}}, \bibinfo {author}
  {\bibfnamefont {B.}~\bibnamefont {Ballesteros}}, \bibinfo {author}
  {\bibfnamefont {A.}~\bibnamefont {Mugarza}}, \bibinfo {author} {\bibfnamefont
  {A.}~\bibnamefont {Barla}}, \bibinfo {author} {\bibfnamefont
  {M.}~\bibnamefont {Valvidares}}, \bibinfo {author} {\bibfnamefont
  {E.}~\bibnamefont {Pellegrin}}, \bibinfo {author} {\bibfnamefont
  {A.}~\bibnamefont {Ghosh}}, \bibinfo {author} {\bibfnamefont {I.~M.}\
  \bibnamefont {Miron}}, \bibinfo {author} {\bibfnamefont {O.}~\bibnamefont
  {Boulle}}, \bibinfo {author} {\bibfnamefont {S.}~\bibnamefont {Auffret}},
  \bibinfo {author} {\bibfnamefont {G.}~\bibnamefont {Gaudin}}, \ and\ \bibinfo
  {author} {\bibfnamefont {P.}~\bibnamefont {Gambardella}},\ }\href {\doibase
  10.1103/PhysRevB.89.214419} {\bibfield  {journal} {\bibinfo  {journal}
  {Physical Review B}\ }\textbf {\bibinfo {volume} {89}},\ \bibinfo {pages}
  {214419} (\bibinfo {year} {2014})}\BibitemShut {NoStop}%
\bibitem [{\citenamefont {Qiu}\ \emph {et~al.}(2015)\citenamefont {Qiu},
  \citenamefont {Narayanapillai}, \citenamefont {Wu}, \citenamefont {Deorani},
  \citenamefont {Yang}, \citenamefont {Noh}, \citenamefont {Park},
  \citenamefont {Lee}, \citenamefont {Lee},\ and\ \citenamefont
  {Yang}}]{Qiu2015}%
  \BibitemOpen
  \bibfield  {author} {\bibinfo {author} {\bibfnamefont {X.}~\bibnamefont
  {Qiu}}, \bibinfo {author} {\bibfnamefont {K.}~\bibnamefont {Narayanapillai}},
  \bibinfo {author} {\bibfnamefont {Y.}~\bibnamefont {Wu}}, \bibinfo {author}
  {\bibfnamefont {P.}~\bibnamefont {Deorani}}, \bibinfo {author} {\bibfnamefont
  {D.-H.}\ \bibnamefont {Yang}}, \bibinfo {author} {\bibfnamefont {W.-s.}\
  \bibnamefont {Noh}}, \bibinfo {author} {\bibfnamefont {J.-h.}\ \bibnamefont
  {Park}}, \bibinfo {author} {\bibfnamefont {K.-J.}\ \bibnamefont {Lee}},
  \bibinfo {author} {\bibfnamefont {H.-w.}\ \bibnamefont {Lee}}, \ and\
  \bibinfo {author} {\bibfnamefont {H.}~\bibnamefont {Yang}},\ }\href {\doibase
  10.1038/nnano.2015.18} {\bibfield  {journal} {\bibinfo  {journal} {Nature
  Nanotechnology}\ }\textbf {\bibinfo {volume} {10}},\ \bibinfo {pages} {333}
  (\bibinfo {year} {2015})}\BibitemShut {NoStop}%
\bibitem [{\citenamefont {Ghosh}\ \emph {et~al.}(2017)\citenamefont {Ghosh},
  \citenamefont {Garello}, \citenamefont {Avci}, \citenamefont {Gabureac},\
  and\ \citenamefont {Gambardella}}]{Ghosh2017}%
  \BibitemOpen
  \bibfield  {author} {\bibinfo {author} {\bibfnamefont {A.}~\bibnamefont
  {Ghosh}}, \bibinfo {author} {\bibfnamefont {K.}~\bibnamefont {Garello}},
  \bibinfo {author} {\bibfnamefont {C.~O.}\ \bibnamefont {Avci}}, \bibinfo
  {author} {\bibfnamefont {M.}~\bibnamefont {Gabureac}}, \ and\ \bibinfo
  {author} {\bibfnamefont {P.}~\bibnamefont {Gambardella}},\ }\href {\doibase
  10.1103/PhysRevApplied.7.014004} {\bibfield  {journal} {\bibinfo  {journal}
  {Physical Review Applied}\ }\textbf {\bibinfo {volume} {7}},\ \bibinfo
  {pages} {014004} (\bibinfo {year} {2017})}\BibitemShut {NoStop}%
\bibitem [{\citenamefont {Liu}\ \emph {et~al.}(2011)\citenamefont {Liu},
  \citenamefont {Moriyama}, \citenamefont {Ralph},\ and\ \citenamefont
  {Buhrman}}]{Liu2011}%
  \BibitemOpen
  \bibfield  {author} {\bibinfo {author} {\bibfnamefont {L.}~\bibnamefont
  {Liu}}, \bibinfo {author} {\bibfnamefont {T.}~\bibnamefont {Moriyama}},
  \bibinfo {author} {\bibfnamefont {D.~C.}\ \bibnamefont {Ralph}}, \ and\
  \bibinfo {author} {\bibfnamefont {R.~A.}\ \bibnamefont {Buhrman}},\ }\href
  {\doibase 10.1103/PhysRevLett.106.036601} {\bibfield  {journal} {\bibinfo
  {journal} {Physical Review Letters}\ }\textbf {\bibinfo {volume} {106}},\
  \bibinfo {pages} {036601} (\bibinfo {year} {2011})}\BibitemShut {NoStop}%
\bibitem [{\citenamefont {Haney}\ \emph
  {et~al.}(2013{\natexlab{a}})\citenamefont {Haney}, \citenamefont {Lee},
  \citenamefont {Lee}, \citenamefont {Manchon},\ and\ \citenamefont
  {Stiles}}]{Haney2013b}%
  \BibitemOpen
  \bibfield  {author} {\bibinfo {author} {\bibfnamefont {P.}~\bibnamefont
  {Haney}}, \bibinfo {author} {\bibfnamefont {H.-W.}\ \bibnamefont {Lee}},
  \bibinfo {author} {\bibfnamefont {K.-J.}\ \bibnamefont {Lee}}, \bibinfo
  {author} {\bibfnamefont {A.}~\bibnamefont {Manchon}}, \ and\ \bibinfo
  {author} {\bibfnamefont {M.}~\bibnamefont {Stiles}},\ }\href {\doibase
  10.1103/PhysRevB.87.174411} {\bibfield  {journal} {\bibinfo  {journal}
  {Physical Review B}\ }\textbf {\bibinfo {volume} {87}},\ \bibinfo {pages}
  {174411} (\bibinfo {year} {2013}{\natexlab{a}})}\BibitemShut {NoStop}%
\bibitem [{\citenamefont {Miron}\ \emph {et~al.}(2010)\citenamefont {Miron},
  \citenamefont {Gaudin}, \citenamefont {Auffret}, \citenamefont {Rodmacq},
  \citenamefont {Schuhl}, \citenamefont {Pizzini}, \citenamefont {Vogel},\ and\
  \citenamefont {Gambardella}}]{Miron2010}%
  \BibitemOpen
  \bibfield  {author} {\bibinfo {author} {\bibfnamefont {I.~M.}\ \bibnamefont
  {Miron}}, \bibinfo {author} {\bibfnamefont {G.}~\bibnamefont {Gaudin}},
  \bibinfo {author} {\bibfnamefont {S.}~\bibnamefont {Auffret}}, \bibinfo
  {author} {\bibfnamefont {B.}~\bibnamefont {Rodmacq}}, \bibinfo {author}
  {\bibfnamefont {A.}~\bibnamefont {Schuhl}}, \bibinfo {author} {\bibfnamefont
  {S.}~\bibnamefont {Pizzini}}, \bibinfo {author} {\bibfnamefont
  {J.}~\bibnamefont {Vogel}}, \ and\ \bibinfo {author} {\bibfnamefont
  {P.}~\bibnamefont {Gambardella}},\ }\href {\doibase 10.1038/nmat2613}
  {\bibfield  {journal} {\bibinfo  {journal} {Nature Materials}\ }\textbf
  {\bibinfo {volume} {9}},\ \bibinfo {pages} {230} (\bibinfo {year}
  {2010})}\BibitemShut {NoStop}%
\bibitem [{\citenamefont {Fan}\ \emph {et~al.}(2013)\citenamefont {Fan},
  \citenamefont {Wu}, \citenamefont {Chen}, \citenamefont {Jerry},
  \citenamefont {Zhang},\ and\ \citenamefont {Xiao}}]{Fan2013}%
  \BibitemOpen
  \bibfield  {author} {\bibinfo {author} {\bibfnamefont {X.}~\bibnamefont
  {Fan}}, \bibinfo {author} {\bibfnamefont {J.}~\bibnamefont {Wu}}, \bibinfo
  {author} {\bibfnamefont {Y.}~\bibnamefont {Chen}}, \bibinfo {author}
  {\bibfnamefont {M.~J.}\ \bibnamefont {Jerry}}, \bibinfo {author}
  {\bibfnamefont {H.}~\bibnamefont {Zhang}}, \ and\ \bibinfo {author}
  {\bibfnamefont {J.~Q.}\ \bibnamefont {Xiao}},\ }\href {\doibase
  10.1038/ncomms2709} {\bibfield  {journal} {\bibinfo  {journal} {Nature
  Communications}\ }\textbf {\bibinfo {volume} {4}},\ \bibinfo {pages} {1799}
  (\bibinfo {year} {2013})}\BibitemShut {NoStop}%
\bibitem [{\citenamefont {Baek}\ \emph {et~al.}(2018)\citenamefont {Baek},
  \citenamefont {Amin}, \citenamefont {Oh}, \citenamefont {Go}, \citenamefont
  {Lee}, \citenamefont {Lee}, \citenamefont {Kim}, \citenamefont {Stiles},
  \citenamefont {Park},\ and\ \citenamefont {Lee}}]{Baek2018}%
  \BibitemOpen
  \bibfield  {author} {\bibinfo {author} {\bibfnamefont {S.~H.~C.}\
  \bibnamefont {Baek}}, \bibinfo {author} {\bibfnamefont {V.~P.}\ \bibnamefont
  {Amin}}, \bibinfo {author} {\bibfnamefont {Y.~W.}\ \bibnamefont {Oh}},
  \bibinfo {author} {\bibfnamefont {G.}~\bibnamefont {Go}}, \bibinfo {author}
  {\bibfnamefont {S.~J.}\ \bibnamefont {Lee}}, \bibinfo {author} {\bibfnamefont
  {G.~H.}\ \bibnamefont {Lee}}, \bibinfo {author} {\bibfnamefont {K.~J.}\
  \bibnamefont {Kim}}, \bibinfo {author} {\bibfnamefont {M.~D.}\ \bibnamefont
  {Stiles}}, \bibinfo {author} {\bibfnamefont {B.~G.}\ \bibnamefont {Park}}, \
  and\ \bibinfo {author} {\bibfnamefont {K.~J.}\ \bibnamefont {Lee}},\ }\href
  {\doibase 10.1038/s41563-018-0041-5} {\bibfield  {journal} {\bibinfo
  {journal} {Nature Materials}\ }\textbf {\bibinfo {volume} {17}},\ \bibinfo
  {pages} {509} (\bibinfo {year} {2018})}\BibitemShut {NoStop}%
\bibitem [{\citenamefont {Safranski}\ \emph {et~al.}(2019)\citenamefont
  {Safranski}, \citenamefont {Montoya},\ and\ \citenamefont
  {Krivorotov}}]{Safranski2019}%
  \BibitemOpen
  \bibfield  {author} {\bibinfo {author} {\bibfnamefont {C.}~\bibnamefont
  {Safranski}}, \bibinfo {author} {\bibfnamefont {E.~A.}\ \bibnamefont
  {Montoya}}, \ and\ \bibinfo {author} {\bibfnamefont {I.~N.}\ \bibnamefont
  {Krivorotov}},\ }\href {\doibase 10.1038/s41565-018-0282-0} {\bibfield
  {journal} {\bibinfo  {journal} {Nature Nanotechnology}\ }\textbf {\bibinfo
  {volume} {14}},\ \bibinfo {pages} {27} (\bibinfo {year} {2019})}\BibitemShut
  {NoStop}%
\bibitem [{\citenamefont {van~der Bijl}\ and\ \citenamefont
  {Duine}(2012)}]{Bijl2012}%
  \BibitemOpen
  \bibfield  {author} {\bibinfo {author} {\bibfnamefont {E.}~\bibnamefont
  {van~der Bijl}}\ and\ \bibinfo {author} {\bibfnamefont {R.~a.}\ \bibnamefont
  {Duine}},\ }\href {\doibase 10.1103/PhysRevB.86.094406} {\bibfield  {journal}
  {\bibinfo  {journal} {Physical Review B}\ }\textbf {\bibinfo {volume} {86}},\
  \bibinfo {pages} {094406} (\bibinfo {year} {2012})}\BibitemShut {NoStop}%
\bibitem [{\citenamefont {Li}\ \emph {et~al.}(2015)\citenamefont {Li},
  \citenamefont {Gao}, \citenamefont {Z{\^{a}}rbo}, \citenamefont
  {V{\'{y}}born{\'{y}}}, \citenamefont {Wang}, \citenamefont {Garate},
  \citenamefont {Dogan}, \citenamefont {Cejchan}, \citenamefont {Sinova},
  \citenamefont {Jungwirth},\ and\ \citenamefont {Manchon}}]{Li2015b}%
  \BibitemOpen
  \bibfield  {author} {\bibinfo {author} {\bibfnamefont {H.}~\bibnamefont
  {Li}}, \bibinfo {author} {\bibfnamefont {H.}~\bibnamefont {Gao}}, \bibinfo
  {author} {\bibfnamefont {L.~P.}\ \bibnamefont {Z{\^{a}}rbo}}, \bibinfo
  {author} {\bibfnamefont {K.}~\bibnamefont {V{\'{y}}born{\'{y}}}}, \bibinfo
  {author} {\bibfnamefont {X.}~\bibnamefont {Wang}}, \bibinfo {author}
  {\bibfnamefont {I.}~\bibnamefont {Garate}}, \bibinfo {author} {\bibfnamefont
  {F.}~\bibnamefont {Dogan}}, \bibinfo {author} {\bibfnamefont
  {A.}~\bibnamefont {Cejchan}}, \bibinfo {author} {\bibfnamefont
  {J.}~\bibnamefont {Sinova}}, \bibinfo {author} {\bibfnamefont
  {T.}~\bibnamefont {Jungwirth}}, \ and\ \bibinfo {author} {\bibfnamefont
  {A.}~\bibnamefont {Manchon}},\ }\href {\doibase 10.1103/PhysRevB.91.134402}
  {\bibfield  {journal} {\bibinfo  {journal} {Physical Review B}\ }\textbf
  {\bibinfo {volume} {91}},\ \bibinfo {pages} {134402} (\bibinfo {year}
  {2015})}\BibitemShut {NoStop}%
\bibitem [{\citenamefont {Qaiumzadeh}\ \emph {et~al.}(2015)\citenamefont
  {Qaiumzadeh}, \citenamefont {Duine},\ and\ \citenamefont
  {Titov}}]{Qaiumzadeh2015}%
  \BibitemOpen
  \bibfield  {author} {\bibinfo {author} {\bibfnamefont {A.}~\bibnamefont
  {Qaiumzadeh}}, \bibinfo {author} {\bibfnamefont {R.~A.}\ \bibnamefont
  {Duine}}, \ and\ \bibinfo {author} {\bibfnamefont {M.}~\bibnamefont
  {Titov}},\ }\href {\doibase 10.1103/PhysRevB.92.014402} {\bibfield  {journal}
  {\bibinfo  {journal} {Physical Review B}\ }\textbf {\bibinfo {volume} {92}},\
  \bibinfo {pages} {014402} (\bibinfo {year} {2015})}\BibitemShut {NoStop}%
\bibitem [{\citenamefont {Ado}\ \emph {et~al.}(2017)\citenamefont {Ado},
  \citenamefont {Tretiakov},\ and\ \citenamefont {Titov}}]{Ado2017}%
  \BibitemOpen
  \bibfield  {author} {\bibinfo {author} {\bibfnamefont {I.~A.}\ \bibnamefont
  {Ado}}, \bibinfo {author} {\bibfnamefont {O.~A.}\ \bibnamefont {Tretiakov}},
  \ and\ \bibinfo {author} {\bibfnamefont {M.}~\bibnamefont {Titov}},\ }\href
  {\doibase 10.1103/PhysRevB.95.094401} {\bibfield  {journal} {\bibinfo
  {journal} {Physical Review B}\ }\textbf {\bibinfo {volume} {95}},\ \bibinfo
  {pages} {094401} (\bibinfo {year} {2017})}\BibitemShut {NoStop}%
\bibitem [{\citenamefont {Chen}\ and\ \citenamefont {Zhang}(2017)}]{Chen2017b}%
  \BibitemOpen
  \bibfield  {author} {\bibinfo {author} {\bibfnamefont {K.}~\bibnamefont
  {Chen}}\ and\ \bibinfo {author} {\bibfnamefont {S.}~\bibnamefont {Zhang}},\
  }\href {\doibase 10.1103/PhysRevB.96.134401} {\bibfield  {journal} {\bibinfo
  {journal} {Physical Review B}\ }\textbf {\bibinfo {volume} {96}},\ \bibinfo
  {pages} {134401} (\bibinfo {year} {2017})}\BibitemShut {NoStop}%
\bibitem [{\citenamefont {Bl{\"{u}}gel}\ and\ \citenamefont
  {Bihlmayer}(2007)}]{Blugel2007}%
  \BibitemOpen
  \bibfield  {author} {\bibinfo {author} {\bibfnamefont {S.}~\bibnamefont
  {Bl{\"{u}}gel}}\ and\ \bibinfo {author} {\bibfnamefont {G.}~\bibnamefont
  {Bihlmayer}},\ }in\ \href {\doibase 10.1002/9780470022184} {\emph {\bibinfo
  {booktitle} {Handbook of Magnetism and Advanced Magnetic Materials.}}},\
  \bibinfo {editor} {edited by\ \bibinfo {editor} {\bibfnamefont
  {H.}~\bibnamefont {Kronmuller}}\ and\ \bibinfo {editor} {\bibfnamefont
  {S.}~\bibnamefont {Parkin}}}\ (\bibinfo  {publisher} {John Wiley {\&} Sons},\
  \bibinfo {year} {2007})\ pp.\ \bibinfo {pages} {1--42}\BibitemShut {NoStop}%
\bibitem [{\citenamefont {Grytsyuk}\ \emph {et~al.}(2016)\citenamefont
  {Grytsyuk}, \citenamefont {Belabbes}, \citenamefont {Haney}, \citenamefont
  {Lee}, \citenamefont {Lee}, \citenamefont {Stiles}, \citenamefont
  {Schwingenschl{\"{o}}gl},\ and\ \citenamefont {Manchon}}]{Grytsyuk2016}%
  \BibitemOpen
  \bibfield  {author} {\bibinfo {author} {\bibfnamefont {S.}~\bibnamefont
  {Grytsyuk}}, \bibinfo {author} {\bibfnamefont {A.}~\bibnamefont {Belabbes}},
  \bibinfo {author} {\bibfnamefont {P.~M.}\ \bibnamefont {Haney}}, \bibinfo
  {author} {\bibfnamefont {H.~W.}\ \bibnamefont {Lee}}, \bibinfo {author}
  {\bibfnamefont {K.~J.}\ \bibnamefont {Lee}}, \bibinfo {author} {\bibfnamefont
  {M.~D.}\ \bibnamefont {Stiles}}, \bibinfo {author} {\bibfnamefont
  {U.}~\bibnamefont {Schwingenschl{\"{o}}gl}}, \ and\ \bibinfo {author}
  {\bibfnamefont {A.}~\bibnamefont {Manchon}},\ }\href {\doibase
  10.1103/PhysRevB.93.174421} {\bibfield  {journal} {\bibinfo  {journal}
  {Physical Review B}\ }\textbf {\bibinfo {volume} {93}},\ \bibinfo {pages}
  {174421} (\bibinfo {year} {2016})}\BibitemShut {NoStop}%
\bibitem [{\citenamefont {Wang}\ \emph
  {et~al.}(2016{\natexlab{a}})\citenamefont {Wang}, \citenamefont {Wesselink},
  \citenamefont {Liu}, \citenamefont {Yuan}, \citenamefont {Xia},\ and\
  \citenamefont {Kelly}}]{Wang2016b}%
  \BibitemOpen
  \bibfield  {author} {\bibinfo {author} {\bibfnamefont {L.}~\bibnamefont
  {Wang}}, \bibinfo {author} {\bibfnamefont {R.~J.~H.}\ \bibnamefont
  {Wesselink}}, \bibinfo {author} {\bibfnamefont {Y.}~\bibnamefont {Liu}},
  \bibinfo {author} {\bibfnamefont {Z.}~\bibnamefont {Yuan}}, \bibinfo {author}
  {\bibfnamefont {K.}~\bibnamefont {Xia}}, \ and\ \bibinfo {author}
  {\bibfnamefont {P.~J.}\ \bibnamefont {Kelly}},\ }\href {\doibase
  10.1103/PhysRevLett.116.196602} {\bibfield  {journal} {\bibinfo  {journal}
  {Physical Review Letters}\ }\textbf {\bibinfo {volume} {116}},\ \bibinfo
  {pages} {196602} (\bibinfo {year} {2016}{\natexlab{a}})}\BibitemShut
  {NoStop}%
\bibitem [{\citenamefont {Tanaka}\ \emph {et~al.}(2008)\citenamefont {Tanaka},
  \citenamefont {Kontani}, \citenamefont {Naito}, \citenamefont {Naito},
  \citenamefont {Hirashima}, \citenamefont {Yamada},\ and\ \citenamefont
  {Inoue}}]{Tanaka2008}%
  \BibitemOpen
  \bibfield  {author} {\bibinfo {author} {\bibfnamefont {T.}~\bibnamefont
  {Tanaka}}, \bibinfo {author} {\bibfnamefont {H.}~\bibnamefont {Kontani}},
  \bibinfo {author} {\bibfnamefont {M.}~\bibnamefont {Naito}}, \bibinfo
  {author} {\bibfnamefont {T.}~\bibnamefont {Naito}}, \bibinfo {author}
  {\bibfnamefont {D.}~\bibnamefont {Hirashima}}, \bibinfo {author}
  {\bibfnamefont {K.}~\bibnamefont {Yamada}}, \ and\ \bibinfo {author}
  {\bibfnamefont {J.}~\bibnamefont {Inoue}},\ }\href {\doibase
  10.1103/PhysRevB.77.165117} {\bibfield  {journal} {\bibinfo  {journal}
  {Physical Review B}\ }\textbf {\bibinfo {volume} {77}},\ \bibinfo {pages}
  {165117} (\bibinfo {year} {2008})}\BibitemShut {NoStop}%
\bibitem [{\citenamefont {Jo}\ \emph {et~al.}(2018)\citenamefont {Jo},
  \citenamefont {Go},\ and\ \citenamefont {Lee}}]{Jo2018}%
  \BibitemOpen
  \bibfield  {author} {\bibinfo {author} {\bibfnamefont {D.}~\bibnamefont
  {Jo}}, \bibinfo {author} {\bibfnamefont {D.}~\bibnamefont {Go}}, \ and\
  \bibinfo {author} {\bibfnamefont {H.-w.}\ \bibnamefont {Lee}},\ }\href
  {\doibase 10.1103/PhysRevB.98.214405} {\bibfield  {journal} {\bibinfo
  {journal} {Physical Review B}\ }\textbf {\bibinfo {volume} {98}},\ \bibinfo
  {pages} {214405} (\bibinfo {year} {2018})}\BibitemShut {NoStop}%
\bibitem [{\citenamefont {Yoda}\ \emph {et~al.}(2018)\citenamefont {Yoda},
  \citenamefont {Yokoyama},\ and\ \citenamefont {Murakami}}]{Yoda2018}%
  \BibitemOpen
  \bibfield  {author} {\bibinfo {author} {\bibfnamefont {T.}~\bibnamefont
  {Yoda}}, \bibinfo {author} {\bibfnamefont {T.}~\bibnamefont {Yokoyama}}, \
  and\ \bibinfo {author} {\bibfnamefont {S.}~\bibnamefont {Murakami}},\ }\href
  {\doibase 10.1021/acs.nanolett.7b04300} {\bibfield  {journal} {\bibinfo
  {journal} {Nano Letters}\ }\textbf {\bibinfo {volume} {18}},\ \bibinfo
  {pages} {916} (\bibinfo {year} {2018})}\BibitemShut {NoStop}%
\bibitem [{\citenamefont {Yao}\ \emph {et~al.}(2004)\citenamefont {Yao},
  \citenamefont {Kleinman}, \citenamefont {MacDonald}, \citenamefont {Sinova},
  \citenamefont {Jungwirth}, \citenamefont {Wang}, \citenamefont {Wang},\ and\
  \citenamefont {Niu}}]{Yao2004}%
  \BibitemOpen
  \bibfield  {author} {\bibinfo {author} {\bibfnamefont {Y.}~\bibnamefont
  {Yao}}, \bibinfo {author} {\bibfnamefont {L.}~\bibnamefont {Kleinman}},
  \bibinfo {author} {\bibfnamefont {A.~H.}\ \bibnamefont {MacDonald}}, \bibinfo
  {author} {\bibfnamefont {J.}~\bibnamefont {Sinova}}, \bibinfo {author}
  {\bibfnamefont {T.}~\bibnamefont {Jungwirth}}, \bibinfo {author}
  {\bibfnamefont {D.-s.}\ \bibnamefont {Wang}}, \bibinfo {author}
  {\bibfnamefont {E.}~\bibnamefont {Wang}}, \ and\ \bibinfo {author}
  {\bibfnamefont {Q.}~\bibnamefont {Niu}},\ }\href {\doibase
  10.1103/PhysRevLett.92.037204} {\bibfield  {journal} {\bibinfo  {journal}
  {Physical Review Letters}\ }\textbf {\bibinfo {volume} {92}},\ \bibinfo
  {pages} {037204} (\bibinfo {year} {2004})}\BibitemShut {NoStop}%
\bibitem [{\citenamefont {Guo}\ \emph {et~al.}(2008)\citenamefont {Guo},
  \citenamefont {Murakami}, \citenamefont {Chen},\ and\ \citenamefont
  {Nagaosa}}]{Guo2008}%
  \BibitemOpen
  \bibfield  {author} {\bibinfo {author} {\bibfnamefont {G.}~\bibnamefont
  {Guo}}, \bibinfo {author} {\bibfnamefont {S.}~\bibnamefont {Murakami}},
  \bibinfo {author} {\bibfnamefont {T.-W.}\ \bibnamefont {Chen}}, \ and\
  \bibinfo {author} {\bibfnamefont {N.}~\bibnamefont {Nagaosa}},\ }\href
  {\doibase 10.1103/PhysRevLett.100.096401} {\bibfield  {journal} {\bibinfo
  {journal} {Physical Review Letters}\ }\textbf {\bibinfo {volume} {100}},\
  \bibinfo {pages} {096401} (\bibinfo {year} {2008})}\BibitemShut {NoStop}%
\bibitem [{\citenamefont {Lowitzer}\ \emph {et~al.}(2011)\citenamefont
  {Lowitzer}, \citenamefont {Gradhand}, \citenamefont {K{\"{o}}dderitzsch},
  \citenamefont {Fedorov}, \citenamefont {Mertig},\ and\ \citenamefont
  {Ebert}}]{Lowitzer2011}%
  \BibitemOpen
  \bibfield  {author} {\bibinfo {author} {\bibfnamefont {S.}~\bibnamefont
  {Lowitzer}}, \bibinfo {author} {\bibfnamefont {M.}~\bibnamefont {Gradhand}},
  \bibinfo {author} {\bibfnamefont {D.}~\bibnamefont {K{\"{o}}dderitzsch}},
  \bibinfo {author} {\bibfnamefont {D.~V.}\ \bibnamefont {Fedorov}}, \bibinfo
  {author} {\bibfnamefont {I.}~\bibnamefont {Mertig}}, \ and\ \bibinfo {author}
  {\bibfnamefont {H.}~\bibnamefont {Ebert}},\ }\href {\doibase
  10.1103/PhysRevLett.106.056601} {\bibfield  {journal} {\bibinfo  {journal}
  {Physical Review Letters}\ }\textbf {\bibinfo {volume} {106}},\ \bibinfo
  {pages} {056601} (\bibinfo {year} {2011})}\BibitemShut {NoStop}%
\bibitem [{\citenamefont {Sun}\ \emph {et~al.}(2016)\citenamefont {Sun},
  \citenamefont {Zhang}, \citenamefont {Felser},\ and\ \citenamefont
  {Yan}}]{Sun2016}%
  \BibitemOpen
  \bibfield  {author} {\bibinfo {author} {\bibfnamefont {Y.}~\bibnamefont
  {Sun}}, \bibinfo {author} {\bibfnamefont {Y.}~\bibnamefont {Zhang}}, \bibinfo
  {author} {\bibfnamefont {C.}~\bibnamefont {Felser}}, \ and\ \bibinfo {author}
  {\bibfnamefont {B.}~\bibnamefont {Yan}},\ }\href {\doibase
  10.1103/PhysRevLett.117.146403} {\bibfield  {journal} {\bibinfo  {journal}
  {Physical Review Letters}\ }\textbf {\bibinfo {volume} {117}},\ \bibinfo
  {pages} {146403} (\bibinfo {year} {2016})},\ \Eprint
  {http://arxiv.org/abs/1604.07167} {arXiv:1604.07167} \BibitemShut {NoStop}%
\bibitem [{\citenamefont {Haney}\ \emph
  {et~al.}(2013{\natexlab{b}})\citenamefont {Haney}, \citenamefont {Lee},
  \citenamefont {Lee}, \citenamefont {Manchon},\ and\ \citenamefont
  {Stiles}}]{Haney2013a}%
  \BibitemOpen
  \bibfield  {author} {\bibinfo {author} {\bibfnamefont {P.~M.}\ \bibnamefont
  {Haney}}, \bibinfo {author} {\bibfnamefont {H.~W.}\ \bibnamefont {Lee}},
  \bibinfo {author} {\bibfnamefont {K.~J.}\ \bibnamefont {Lee}}, \bibinfo
  {author} {\bibfnamefont {A.}~\bibnamefont {Manchon}}, \ and\ \bibinfo
  {author} {\bibfnamefont {M.~D.}\ \bibnamefont {Stiles}},\ }\href@noop {}
  {\bibfield  {journal} {\bibinfo  {journal} {Physical Review B}\ }\textbf
  {\bibinfo {volume} {88}},\ \bibinfo {pages} {214417} (\bibinfo {year}
  {2013}{\natexlab{b}})}\BibitemShut {NoStop}%
\bibitem [{\citenamefont {Freimuth}\ \emph {et~al.}(2014)\citenamefont
  {Freimuth}, \citenamefont {Bl{\"{u}}gel},\ and\ \citenamefont
  {Mokrousov}}]{Freimuth2014a}%
  \BibitemOpen
  \bibfield  {author} {\bibinfo {author} {\bibfnamefont {F.}~\bibnamefont
  {Freimuth}}, \bibinfo {author} {\bibfnamefont {S.}~\bibnamefont
  {Bl{\"{u}}gel}}, \ and\ \bibinfo {author} {\bibfnamefont {Y.}~\bibnamefont
  {Mokrousov}},\ }\href {\doibase 10.1103/PhysRevB.90.174423} {\bibfield
  {journal} {\bibinfo  {journal} {Physical Review B}\ }\textbf {\bibinfo
  {volume} {90}},\ \bibinfo {pages} {174423} (\bibinfo {year}
  {2014})}\BibitemShut {NoStop}%
\bibitem [{\citenamefont {G{\'{e}}ranton}\ \emph {et~al.}(2015)\citenamefont
  {G{\'{e}}ranton}, \citenamefont {Freimuth}, \citenamefont {Bl{\"{u}}gel},\
  and\ \citenamefont {Mokrousov}}]{Geranton2015}%
  \BibitemOpen
  \bibfield  {author} {\bibinfo {author} {\bibfnamefont {G.}~\bibnamefont
  {G{\'{e}}ranton}}, \bibinfo {author} {\bibfnamefont {F.}~\bibnamefont
  {Freimuth}}, \bibinfo {author} {\bibfnamefont {S.}~\bibnamefont
  {Bl{\"{u}}gel}}, \ and\ \bibinfo {author} {\bibfnamefont {Y.}~\bibnamefont
  {Mokrousov}},\ }\href {\doibase 10.1103/PhysRevB.91.014417} {\bibfield
  {journal} {\bibinfo  {journal} {Physical Review B}\ }\textbf {\bibinfo
  {volume} {91}},\ \bibinfo {pages} {014417} (\bibinfo {year}
  {2015})}\BibitemShut {NoStop}%
\bibitem [{\citenamefont {G{\'{e}}ranton}\ \emph {et~al.}(2016)\citenamefont
  {G{\'{e}}ranton}, \citenamefont {Zimmermann}, \citenamefont {Long},
  \citenamefont {Mavropoulos}, \citenamefont {Bl{\"{u}}gel}, \citenamefont
  {Freimuth},\ and\ \citenamefont {Mokrousov}}]{Geranton2016}%
  \BibitemOpen
  \bibfield  {author} {\bibinfo {author} {\bibfnamefont {G.}~\bibnamefont
  {G{\'{e}}ranton}}, \bibinfo {author} {\bibfnamefont {B.}~\bibnamefont
  {Zimmermann}}, \bibinfo {author} {\bibfnamefont {N.~H.}\ \bibnamefont
  {Long}}, \bibinfo {author} {\bibfnamefont {P.}~\bibnamefont {Mavropoulos}},
  \bibinfo {author} {\bibfnamefont {S.}~\bibnamefont {Bl{\"{u}}gel}}, \bibinfo
  {author} {\bibfnamefont {F.}~\bibnamefont {Freimuth}}, \ and\ \bibinfo
  {author} {\bibfnamefont {Y.}~\bibnamefont {Mokrousov}},\ }\href {\doibase
  10.1103/PhysRevB.93.224420} {\bibfield  {journal} {\bibinfo  {journal}
  {Physical Review B}\ }\textbf {\bibinfo {volume} {93}},\ \bibinfo {pages}
  {224420} (\bibinfo {year} {2016})}\BibitemShut {NoStop}%
\bibitem [{\citenamefont {Mahfouzi}\ and\ \citenamefont
  {Kioussis}(2018)}]{Mahfouzi2018a}%
  \BibitemOpen
  \bibfield  {author} {\bibinfo {author} {\bibfnamefont {F.}~\bibnamefont
  {Mahfouzi}}\ and\ \bibinfo {author} {\bibfnamefont {N.}~\bibnamefont
  {Kioussis}},\ }\href {\doibase 10.1103/PhysRevB.97.224426} {\bibfield
  {journal} {\bibinfo  {journal} {Physical Review B}\ }\textbf {\bibinfo
  {volume} {97}},\ \bibinfo {pages} {224426} (\bibinfo {year}
  {2018})}\BibitemShut {NoStop}%
\bibitem [{\citenamefont {Wimmer}\ \emph {et~al.}(2016)\citenamefont {Wimmer},
  \citenamefont {Chadova}, \citenamefont {Seemann}, \citenamefont
  {K{\"{o}}dderitzsch},\ and\ \citenamefont {Ebert}}]{Wimmer2016}%
  \BibitemOpen
  \bibfield  {author} {\bibinfo {author} {\bibfnamefont {S.}~\bibnamefont
  {Wimmer}}, \bibinfo {author} {\bibfnamefont {K.}~\bibnamefont {Chadova}},
  \bibinfo {author} {\bibfnamefont {M.}~\bibnamefont {Seemann}}, \bibinfo
  {author} {\bibfnamefont {D.}~\bibnamefont {K{\"{o}}dderitzsch}}, \ and\
  \bibinfo {author} {\bibfnamefont {H.}~\bibnamefont {Ebert}},\ }\href
  {\doibase 10.1103/PhysRevB.94.054415} {\bibfield  {journal} {\bibinfo
  {journal} {Physical Review B}\ }\textbf {\bibinfo {volume} {94}},\ \bibinfo
  {pages} {054415} (\bibinfo {year} {2016})}\BibitemShut {NoStop}%
\bibitem [{\citenamefont {Ebert}\ \emph {et~al.}(2011)\citenamefont {Ebert},
  \citenamefont {K{\"{o}}dderitzsch},\ and\ \citenamefont
  {Min{\'{a}}r}}]{Ebert2011b}%
  \BibitemOpen
  \bibfield  {author} {\bibinfo {author} {\bibfnamefont {H.}~\bibnamefont
  {Ebert}}, \bibinfo {author} {\bibfnamefont {D.}~\bibnamefont
  {K{\"{o}}dderitzsch}}, \ and\ \bibinfo {author} {\bibfnamefont
  {J.}~\bibnamefont {Min{\'{a}}r}},\ }\href {\doibase
  10.1088/0034-4885/74/9/096501} {\bibfield  {journal} {\bibinfo  {journal}
  {Reports on Progress in Physics}\ }\textbf {\bibinfo {volume} {74}},\
  \bibinfo {pages} {096501} (\bibinfo {year} {2011})}\BibitemShut {NoStop}%
\bibitem [{\citenamefont {Wang}\ \emph
  {et~al.}(2016{\natexlab{b}})\citenamefont {Wang}, \citenamefont
  {Chotorlishvili}, \citenamefont {Guo}, \citenamefont {Sukhov}, \citenamefont
  {Dugaev}, \citenamefont {Barnas},\ and\ \citenamefont {Berakdar}}]{Wang2016}%
  \BibitemOpen
  \bibfield  {author} {\bibinfo {author} {\bibfnamefont {X.~G.}\ \bibnamefont
  {Wang}}, \bibinfo {author} {\bibfnamefont {L.}~\bibnamefont
  {Chotorlishvili}}, \bibinfo {author} {\bibfnamefont {G.~H.}\ \bibnamefont
  {Guo}}, \bibinfo {author} {\bibfnamefont {A.}~\bibnamefont {Sukhov}},
  \bibinfo {author} {\bibfnamefont {V.}~\bibnamefont {Dugaev}}, \bibinfo
  {author} {\bibfnamefont {J.}~\bibnamefont {Barnas}}, \ and\ \bibinfo {author}
  {\bibfnamefont {J.}~\bibnamefont {Berakdar}},\ }\href {\doibase
  10.1103/PhysRevB.94.104410} {\bibfield  {journal} {\bibinfo  {journal}
  {Physical Review B}\ }\textbf {\bibinfo {volume} {94}},\ \bibinfo {pages}
  {104410} (\bibinfo {year} {2016}{\natexlab{b}})}\BibitemShut {NoStop}%
\bibitem [{\citenamefont {Belashchenko}\ \emph {et~al.}(2019)\citenamefont
  {Belashchenko}, \citenamefont {Kovalev},\ and\ \citenamefont
  {Schilfgaarde}}]{Belashchenko2019}%
  \BibitemOpen
  \bibfield  {author} {\bibinfo {author} {\bibfnamefont {K.~D.}\ \bibnamefont
  {Belashchenko}}, \bibinfo {author} {\bibfnamefont {A.~A.}\ \bibnamefont
  {Kovalev}}, \ and\ \bibinfo {author} {\bibfnamefont {M.~V.}\ \bibnamefont
  {Schilfgaarde}},\ }\href {\doibase 10.1103/PhysRevMaterials.3.011401}
  {\bibfield  {journal} {\bibinfo  {journal} {Physical Review Materials}\
  }\textbf {\bibinfo {volume} {3}},\ \bibinfo {pages} {011401(R)} (\bibinfo
  {year} {2019})}\BibitemShut {NoStop}%
\bibitem [{\citenamefont {Papaconstantopoulos}\ and\ \citenamefont
  {Mehl}(2003)}]{Papaconstantopoulos2003}%
  \BibitemOpen
  \bibfield  {author} {\bibinfo {author} {\bibfnamefont {D.~a.}\ \bibnamefont
  {Papaconstantopoulos}}\ and\ \bibinfo {author} {\bibfnamefont {M.~J.}\
  \bibnamefont {Mehl}},\ }\href {\doibase 10.1088/0953-8984/15/10/201}
  {\bibfield  {journal} {\bibinfo  {journal} {Journal of Physics: Condensed
  Matter}\ }\textbf {\bibinfo {volume} {15}},\ \bibinfo {pages} {R413}
  (\bibinfo {year} {2003})}\BibitemShut {NoStop}%
\bibitem [{\citenamefont
  {Papaconstantopoulos}(2015)}]{Papaconstantopoulos2015}%
  \BibitemOpen
  \bibfield  {author} {\bibinfo {author} {\bibfnamefont {D.~A.}\ \bibnamefont
  {Papaconstantopoulos}},\ }\href {\doibase 10.1016/0001-6160(66)90168-4}
  {\emph {\bibinfo {title} {{Handbook of the Band Structure of Elemental
  Solids}}}},\ \bibinfo {edition} {second edi}\ ed.\ (\bibinfo  {publisher}
  {Springer, New York},\ \bibinfo {year} {2015})\ p.\ \bibinfo {pages}
  {410}\BibitemShut {NoStop}%
\bibitem [{\citenamefont {Barreteau}\ \emph {et~al.}(2016)\citenamefont
  {Barreteau}, \citenamefont {Spanjaard},\ and\ \citenamefont
  {Desjonqu{\`{e}}res}}]{Barreteau2016}%
  \BibitemOpen
  \bibfield  {author} {\bibinfo {author} {\bibfnamefont {C.}~\bibnamefont
  {Barreteau}}, \bibinfo {author} {\bibfnamefont {D.}~\bibnamefont
  {Spanjaard}}, \ and\ \bibinfo {author} {\bibfnamefont {M.-c.}\ \bibnamefont
  {Desjonqu{\`{e}}res}},\ }\href {\doibase 10.1016/j.crhy.2015.12.014}
  {\bibfield  {journal} {\bibinfo  {journal} {Comptes Rendus Physique}\
  }\textbf {\bibinfo {volume} {17}},\ \bibinfo {pages} {406} (\bibinfo {year}
  {2016})}\BibitemShut {NoStop}%
\bibitem [{\citenamefont {Guo}\ \emph {et~al.}(2005)\citenamefont {Guo},
  \citenamefont {Yao},\ and\ \citenamefont {Niu}}]{Guo2005}%
  \BibitemOpen
  \bibfield  {author} {\bibinfo {author} {\bibfnamefont {G.~Y.}\ \bibnamefont
  {Guo}}, \bibinfo {author} {\bibfnamefont {Y.}~\bibnamefont {Yao}}, \ and\
  \bibinfo {author} {\bibfnamefont {Q.}~\bibnamefont {Niu}},\ }\href {\doibase
  10.1103/PhysRevLett.94.226601} {\bibfield  {journal} {\bibinfo  {journal}
  {Physical Review Letters}\ }\textbf {\bibinfo {volume} {94}},\ \bibinfo
  {pages} {226601} (\bibinfo {year} {2005})}\BibitemShut {NoStop}%
\bibitem [{\citenamefont {Yao}\ and\ \citenamefont {Fang}(2005)}]{Yao2005}%
  \BibitemOpen
  \bibfield  {author} {\bibinfo {author} {\bibfnamefont {Y.}~\bibnamefont
  {Yao}}\ and\ \bibinfo {author} {\bibfnamefont {Z.}~\bibnamefont {Fang}},\
  }\href {\doibase 10.1103/PhysRevLett.95.156601} {\bibfield  {journal}
  {\bibinfo  {journal} {Physical Review Letters}\ }\textbf {\bibinfo {volume}
  {95}},\ \bibinfo {pages} {156601} (\bibinfo {year} {2005})},\ \Eprint
  {http://arxiv.org/abs/0502351} {arXiv:0502351 [cond-mat]} \BibitemShut
  {NoStop}%
\bibitem [{\citenamefont {Freimuth}\ \emph {et~al.}(2010)\citenamefont
  {Freimuth}, \citenamefont {Bl{\"{u}}gel},\ and\ \citenamefont
  {Mokrousov}}]{Freimuth2010}%
  \BibitemOpen
  \bibfield  {author} {\bibinfo {author} {\bibfnamefont {F.}~\bibnamefont
  {Freimuth}}, \bibinfo {author} {\bibfnamefont {S.}~\bibnamefont
  {Bl{\"{u}}gel}}, \ and\ \bibinfo {author} {\bibfnamefont {Y.}~\bibnamefont
  {Mokrousov}},\ }\href {\doibase 10.1103/PhysRevLett.105.246602} {\bibfield
  {journal} {\bibinfo  {journal} {Physical Review Letters}\ }\textbf {\bibinfo
  {volume} {105}},\ \bibinfo {pages} {246602} (\bibinfo {year}
  {2010})}\BibitemShut {NoStop}%
\bibitem [{\citenamefont {Şahin}\ and\ \citenamefont
  {Flatt{\'{e}}}(2015)}]{Sahin2015}%
  \BibitemOpen
  \bibfield  {author} {\bibinfo {author} {\bibfnamefont {C.}~\bibnamefont
  {Şahin}}\ and\ \bibinfo {author} {\bibfnamefont {M.~E.}\ \bibnamefont
  {Flatt{\'{e}}}},\ }\href {\doibase 10.1103/PhysRevLett.114.107201} {\bibfield
   {journal} {\bibinfo  {journal} {Physical Review Letters}\ }\textbf {\bibinfo
  {volume} {114}},\ \bibinfo {pages} {107201} (\bibinfo {year}
  {2015})}\BibitemShut {NoStop}%
\bibitem [{\citenamefont {Ghosh}\ and\ \citenamefont
  {Manchon}(2018)}]{Ghosh2018}%
  \BibitemOpen
  \bibfield  {author} {\bibinfo {author} {\bibfnamefont {S.}~\bibnamefont
  {Ghosh}}\ and\ \bibinfo {author} {\bibfnamefont {A.}~\bibnamefont
  {Manchon}},\ }\href {\doibase 10.1103/PhysRevB.97.134402} {\bibfield
  {journal} {\bibinfo  {journal} {Physical Review B}\ }\textbf {\bibinfo
  {volume} {97}},\ \bibinfo {pages} {134402} (\bibinfo {year}
  {2018})}\BibitemShut {NoStop}%
\bibitem [{\citenamefont {Ghosh}\ and\ \citenamefont
  {Manchon}(2019)}]{Ghosh2019}%
  \BibitemOpen
  \bibfield  {author} {\bibinfo {author} {\bibfnamefont {S.}~\bibnamefont
  {Ghosh}}\ and\ \bibinfo {author} {\bibfnamefont {A.}~\bibnamefont
  {Manchon}},\ }\href {\doibase 10.1103/PhysRevB.100.014412} {\bibfield
  {journal} {\bibinfo  {journal} {Physical Review B}\ }\textbf {\bibinfo
  {volume} {100}},\ \bibinfo {pages} {014412} (\bibinfo {year}
  {2019})}\BibitemShut {NoStop}%
\bibitem [{\citenamefont {Marchand}\ and\ \citenamefont
  {Franz}(2012)}]{Marchand2012}%
  \BibitemOpen
  \bibfield  {author} {\bibinfo {author} {\bibfnamefont {D.~J.~J.}\
  \bibnamefont {Marchand}}\ and\ \bibinfo {author} {\bibfnamefont
  {M.}~\bibnamefont {Franz}},\ }\href {\doibase 10.1103/PhysRevB.86.155146}
  {\bibfield  {journal} {\bibinfo  {journal} {Physical Review B}\ }\textbf
  {\bibinfo {volume} {86}},\ \bibinfo {pages} {155146} (\bibinfo {year}
  {2012})}\BibitemShut {NoStop}%
\bibitem [{\citenamefont {Pauyac}\ \emph {et~al.}(2018)\citenamefont {Pauyac},
  \citenamefont {Chshiev}, \citenamefont {Manchon},\ and\ \citenamefont
  {Nikolaev}}]{Pauyac2018}%
  \BibitemOpen
  \bibfield  {author} {\bibinfo {author} {\bibfnamefont {C.~O.}\ \bibnamefont
  {Pauyac}}, \bibinfo {author} {\bibfnamefont {M.}~\bibnamefont {Chshiev}},
  \bibinfo {author} {\bibfnamefont {A.}~\bibnamefont {Manchon}}, \ and\
  \bibinfo {author} {\bibfnamefont {S.~A.}\ \bibnamefont {Nikolaev}},\ }\href
  {\doibase 10.1103/PhysRevLett.120.176802} {\bibfield  {journal} {\bibinfo
  {journal} {Physical Review Letters}\ }\textbf {\bibinfo {volume} {120}},\
  \bibinfo {pages} {176802} (\bibinfo {year} {2018})}\BibitemShut {NoStop}%
\bibitem [{\citenamefont {Wang}\ \emph {et~al.}(2019)\citenamefont {Wang},
  \citenamefont {Wang}, \citenamefont {Amin}, \citenamefont {Wang},
  \citenamefont {Radhakrishnan}, \citenamefont {Davidson}, \citenamefont
  {Allen}, \citenamefont {Silva}, \citenamefont {Ohldag}, \citenamefont
  {Balzar}, \citenamefont {Zink}, \citenamefont {Haney}, \citenamefont {Xiao},
  \citenamefont {Cahill}, \citenamefont {Lorenz},\ and\ \citenamefont
  {Fan}}]{Wang2019b}%
  \BibitemOpen
  \bibfield  {author} {\bibinfo {author} {\bibfnamefont {W.}~\bibnamefont
  {Wang}}, \bibinfo {author} {\bibfnamefont {T.}~\bibnamefont {Wang}}, \bibinfo
  {author} {\bibfnamefont {V.~P.}\ \bibnamefont {Amin}}, \bibinfo {author}
  {\bibfnamefont {Y.}~\bibnamefont {Wang}}, \bibinfo {author} {\bibfnamefont
  {A.}~\bibnamefont {Radhakrishnan}}, \bibinfo {author} {\bibfnamefont
  {A.}~\bibnamefont {Davidson}}, \bibinfo {author} {\bibfnamefont {S.~R.}\
  \bibnamefont {Allen}}, \bibinfo {author} {\bibfnamefont {T.~J.}\ \bibnamefont
  {Silva}}, \bibinfo {author} {\bibfnamefont {H.}~\bibnamefont {Ohldag}},
  \bibinfo {author} {\bibfnamefont {D.}~\bibnamefont {Balzar}}, \bibinfo
  {author} {\bibfnamefont {B.~L.}\ \bibnamefont {Zink}}, \bibinfo {author}
  {\bibfnamefont {P.~M.}\ \bibnamefont {Haney}}, \bibinfo {author}
  {\bibfnamefont {J.~Q.}\ \bibnamefont {Xiao}}, \bibinfo {author}
  {\bibfnamefont {D.~G.}\ \bibnamefont {Cahill}}, \bibinfo {author}
  {\bibfnamefont {V.~O.}\ \bibnamefont {Lorenz}}, \ and\ \bibinfo {author}
  {\bibfnamefont {X.}~\bibnamefont {Fan}},\ }\href {\doibase
  10.1038/s41565-019-0504-0} {\bibfield  {journal} {\bibinfo  {journal} {Nature
  Nanotechnology}\ }\textbf {\bibinfo {volume} {14}},\ \bibinfo {pages} {819}
  (\bibinfo {year} {2019})}\BibitemShut {NoStop}%
\bibitem [{\citenamefont {Murakami}\ \emph {et~al.}(2003)\citenamefont
  {Murakami}, \citenamefont {Nagaosa},\ and\ \citenamefont
  {Zhang}}]{Murakami2003}%
  \BibitemOpen
  \bibfield  {author} {\bibinfo {author} {\bibfnamefont {S.}~\bibnamefont
  {Murakami}}, \bibinfo {author} {\bibfnamefont {N.}~\bibnamefont {Nagaosa}}, \
  and\ \bibinfo {author} {\bibfnamefont {S.-C.}\ \bibnamefont {Zhang}},\
  }\href@noop {} {\bibfield  {journal} {\bibinfo  {journal} {Science (New York,
  N.Y.)}\ }\textbf {\bibinfo {volume} {301}},\ \bibinfo {pages} {1348}
  (\bibinfo {year} {2003})}\BibitemShut {NoStop}%
\bibitem [{\citenamefont {Sinova}\ \emph {et~al.}(2004)\citenamefont {Sinova},
  \citenamefont {Culcer}, \citenamefont {Niu}, \citenamefont {Sinitsyn},
  \citenamefont {Jungwirth},\ and\ \citenamefont {MacDonald}}]{Sinova2004}%
  \BibitemOpen
  \bibfield  {author} {\bibinfo {author} {\bibfnamefont {J.}~\bibnamefont
  {Sinova}}, \bibinfo {author} {\bibfnamefont {D.}~\bibnamefont {Culcer}},
  \bibinfo {author} {\bibfnamefont {Q.}~\bibnamefont {Niu}}, \bibinfo {author}
  {\bibfnamefont {N.~A.}\ \bibnamefont {Sinitsyn}}, \bibinfo {author}
  {\bibfnamefont {T.}~\bibnamefont {Jungwirth}}, \ and\ \bibinfo {author}
  {\bibfnamefont {A.~H.}\ \bibnamefont {MacDonald}},\ }\href {\doibase
  10.1103/PhysRevLett.92.126603} {\bibfield  {journal} {\bibinfo  {journal}
  {Physical Review Letters}\ }\textbf {\bibinfo {volume} {92}},\ \bibinfo
  {pages} {126603} (\bibinfo {year} {2004})}\BibitemShut {NoStop}%
\bibitem [{\citenamefont {Kontani}\ \emph {et~al.}(2009)\citenamefont
  {Kontani}, \citenamefont {Tanaka}, \citenamefont {Hirashima}, \citenamefont
  {Yamada},\ and\ \citenamefont {Inoue}}]{Kontani2009}%
  \BibitemOpen
  \bibfield  {author} {\bibinfo {author} {\bibfnamefont {H.}~\bibnamefont
  {Kontani}}, \bibinfo {author} {\bibfnamefont {T.}~\bibnamefont {Tanaka}},
  \bibinfo {author} {\bibfnamefont {D.}~\bibnamefont {Hirashima}}, \bibinfo
  {author} {\bibfnamefont {K.}~\bibnamefont {Yamada}}, \ and\ \bibinfo {author}
  {\bibfnamefont {J.}~\bibnamefont {Inoue}},\ }\href {\doibase
  10.1103/PhysRevLett.102.016601} {\bibfield  {journal} {\bibinfo  {journal}
  {Phys. Rev. Lett.}\ }\textbf {\bibinfo {volume} {102}},\ \bibinfo {pages}
  {016601} (\bibinfo {year} {2009})}\BibitemShut {NoStop}%
\bibitem [{\citenamefont {{F.T. Vasko}}(1979)}]{Vasko1979}%
  \BibitemOpen
  \bibfield  {author} {\bibinfo {author} {\bibnamefont {{F.T. Vasko}}},\
  }\href@noop {} {\bibfield  {journal} {\bibinfo  {journal} {Pis'ma Zh. Eksp.
  Teor. Fiz}\ }\textbf {\bibinfo {volume} {30}},\ \bibinfo {pages} {574}
  (\bibinfo {year} {1979})}\BibitemShut {NoStop}%
\bibitem [{\citenamefont {Ohkawa}\ and\ \citenamefont
  {Uemura}(1974)}]{Ohkawa1974}%
  \BibitemOpen
  \bibfield  {author} {\bibinfo {author} {\bibfnamefont {F.~J.}\ \bibnamefont
  {Ohkawa}}\ and\ \bibinfo {author} {\bibfnamefont {Y.}~\bibnamefont
  {Uemura}},\ }\href {\doibase 10.1143/JPSJ.37.1325} {\bibfield  {journal}
  {\bibinfo  {journal} {Journal of the Physical Society of Japan}\ }\textbf
  {\bibinfo {volume} {37}},\ \bibinfo {pages} {1325} (\bibinfo {year}
  {1974})}\BibitemShut {NoStop}%
\bibitem [{\citenamefont {Bychkov}\ and\ \citenamefont
  {Rashba}(1984)}]{Bychkov1984}%
  \BibitemOpen
  \bibfield  {author} {\bibinfo {author} {\bibfnamefont {Y.~a.}\ \bibnamefont
  {Bychkov}}\ and\ \bibinfo {author} {\bibfnamefont {E.~I.}\ \bibnamefont
  {Rashba}},\ }\href {\doibase 10.1088/0022-3719/17/33/015} {\bibfield
  {journal} {\bibinfo  {journal} {Journal of Physics C: Solid State Physics}\
  }\textbf {\bibinfo {volume} {17}},\ \bibinfo {pages} {6039} (\bibinfo {year}
  {1984})}\BibitemShut {NoStop}%
\bibitem [{\citenamefont {Petersen}\ and\ \citenamefont
  {Hedegard}(2000)}]{Petersen2000}%
  \BibitemOpen
  \bibfield  {author} {\bibinfo {author} {\bibfnamefont {L.}~\bibnamefont
  {Petersen}}\ and\ \bibinfo {author} {\bibfnamefont {P.}~\bibnamefont
  {Hedegard}},\ }\href@noop {} {\bibfield  {journal} {\bibinfo  {journal}
  {Surface Science}\ }\textbf {\bibinfo {volume} {459}},\ \bibinfo {pages} {49}
  (\bibinfo {year} {2000})}\BibitemShut {NoStop}%
\bibitem [{\citenamefont {Bihlmayer}\ \emph {et~al.}(2006)\citenamefont
  {Bihlmayer}, \citenamefont {Koroteev}, \citenamefont {Echenique},
  \citenamefont {Chulkov},\ and\ \citenamefont {Bl{\"{u}}gel}}]{Bihlmayer2006}%
  \BibitemOpen
  \bibfield  {author} {\bibinfo {author} {\bibfnamefont {G.}~\bibnamefont
  {Bihlmayer}}, \bibinfo {author} {\bibfnamefont {Y.}~\bibnamefont {Koroteev}},
  \bibinfo {author} {\bibfnamefont {P.}~\bibnamefont {Echenique}}, \bibinfo
  {author} {\bibfnamefont {E.}~\bibnamefont {Chulkov}}, \ and\ \bibinfo
  {author} {\bibfnamefont {S.}~\bibnamefont {Bl{\"{u}}gel}},\ }\href {\doibase
  10.1016/j.susc.2006.01.098} {\bibfield  {journal} {\bibinfo  {journal}
  {Surface Science}\ }\textbf {\bibinfo {volume} {600}},\ \bibinfo {pages}
  {3888} (\bibinfo {year} {2006})}\BibitemShut {NoStop}%
\bibitem [{\citenamefont {Slater}\ and\ \citenamefont
  {Koster}(1954)}]{Slater1954}%
  \BibitemOpen
  \bibfield  {author} {\bibinfo {author} {\bibfnamefont {J.~C.}\ \bibnamefont
  {Slater}}\ and\ \bibinfo {author} {\bibfnamefont {G.~F.}\ \bibnamefont
  {Koster}},\ }\href {\doibase 10.1103/PhysRev.94.1498} {\bibfield  {journal}
  {\bibinfo  {journal} {Physical Review}\ }\textbf {\bibinfo {volume} {94}},\
  \bibinfo {pages} {1498} (\bibinfo {year} {1954})}\BibitemShut {NoStop}%
\bibitem [{\citenamefont {Manchon}\ \emph {et~al.}(2015)\citenamefont
  {Manchon}, \citenamefont {Koo}, \citenamefont {Nitta}, \citenamefont
  {Frolov},\ and\ \citenamefont {Duine}}]{Manchon2015}%
  \BibitemOpen
  \bibfield  {author} {\bibinfo {author} {\bibfnamefont {A.}~\bibnamefont
  {Manchon}}, \bibinfo {author} {\bibfnamefont {H.~C.}\ \bibnamefont {Koo}},
  \bibinfo {author} {\bibfnamefont {J.}~\bibnamefont {Nitta}}, \bibinfo
  {author} {\bibfnamefont {S.~M.}\ \bibnamefont {Frolov}}, \ and\ \bibinfo
  {author} {\bibfnamefont {R.~A.}\ \bibnamefont {Duine}},\ }\href {\doibase
  10.1038/nmat4360} {\bibfield  {journal} {\bibinfo  {journal} {Nature
  Materials}\ }\textbf {\bibinfo {volume} {14}},\ \bibinfo {pages} {871}
  (\bibinfo {year} {2015})}\BibitemShut {NoStop}%
\bibitem [{\citenamefont {Kresse}\ and\ \citenamefont
  {Furthmuller}(1996)}]{Kresse1996a}%
  \BibitemOpen
  \bibfield  {author} {\bibinfo {author} {\bibfnamefont {G.}~\bibnamefont
  {Kresse}}\ and\ \bibinfo {author} {\bibfnamefont {J.}~\bibnamefont
  {Furthmuller}},\ }\href@noop {} {\bibfield  {journal} {\bibinfo  {journal}
  {Physical Review B}\ }\textbf {\bibinfo {volume} {54}},\ \bibinfo {pages}
  {11169} (\bibinfo {year} {1996})}\BibitemShut {NoStop}%
\bibitem [{\citenamefont {Kresse}\ and\ \citenamefont
  {Furthmiiller}(1996)}]{Kresse1996b}%
  \BibitemOpen
  \bibfield  {author} {\bibinfo {author} {\bibfnamefont {G.}~\bibnamefont
  {Kresse}}\ and\ \bibinfo {author} {\bibfnamefont {J.}~\bibnamefont
  {Furthmiiller}},\ }\href@noop {} {\bibfield  {journal} {\bibinfo  {journal}
  {Computational Materials Science}\ }\textbf {\bibinfo {volume} {6}},\
  \bibinfo {pages} {15} (\bibinfo {year} {1996})}\BibitemShut {NoStop}%
\bibitem [{\citenamefont {Blochl}(1994)}]{Blochl1994}%
  \BibitemOpen
  \bibfield  {author} {\bibinfo {author} {\bibfnamefont {P.~E.}\ \bibnamefont
  {Blochl}},\ }\href@noop {} {\bibfield  {journal} {\bibinfo  {journal}
  {Physical Review B}\ }\textbf {\bibinfo {volume} {50}},\ \bibinfo {pages}
  {17953} (\bibinfo {year} {1994})}\BibitemShut {NoStop}%
\bibitem [{\citenamefont {Kresse}\ and\ \citenamefont
  {Joubert}(1999)}]{Kresse1999}%
  \BibitemOpen
  \bibfield  {author} {\bibinfo {author} {\bibfnamefont {G.}~\bibnamefont
  {Kresse}}\ and\ \bibinfo {author} {\bibfnamefont {D.}~\bibnamefont
  {Joubert}},\ }\href@noop {} {\bibfield  {journal} {\bibinfo  {journal}
  {Physical Review B}\ }\textbf {\bibinfo {volume} {59}},\ \bibinfo {pages}
  {1758} (\bibinfo {year} {1999})}\BibitemShut {NoStop}%
\bibitem [{\citenamefont {Sinitsyn}\ \emph {et~al.}(2006)\citenamefont
  {Sinitsyn}, \citenamefont {Hill}, \citenamefont {Min}, \citenamefont
  {Sinova},\ and\ \citenamefont {MacDonald}}]{Sinitsyn2006}%
  \BibitemOpen
  \bibfield  {author} {\bibinfo {author} {\bibfnamefont {N.~A.}\ \bibnamefont
  {Sinitsyn}}, \bibinfo {author} {\bibfnamefont {J.~E.}\ \bibnamefont {Hill}},
  \bibinfo {author} {\bibfnamefont {H.}~\bibnamefont {Min}}, \bibinfo {author}
  {\bibfnamefont {J.}~\bibnamefont {Sinova}}, \ and\ \bibinfo {author}
  {\bibfnamefont {A.~H.}\ \bibnamefont {MacDonald}},\ }\href {\doibase
  10.1103/PhysRevLett.97.106804} {\bibfield  {journal} {\bibinfo  {journal}
  {Physical Review Letters}\ }\textbf {\bibinfo {volume} {97}},\ \bibinfo
  {pages} {106804} (\bibinfo {year} {2006})},\ \Eprint
  {http://arxiv.org/abs/0602598} {arXiv:0602598 [cond-mat]} \BibitemShut
  {NoStop}%
\bibitem [{\citenamefont {Fan}\ \emph {et~al.}(2014)\citenamefont {Fan},
  \citenamefont {Celik}, \citenamefont {Wu}, \citenamefont {Ni}, \citenamefont
  {Lee}, \citenamefont {Lorenz},\ and\ \citenamefont {Xiao}}]{Fan2014c}%
  \BibitemOpen
  \bibfield  {author} {\bibinfo {author} {\bibfnamefont {X.}~\bibnamefont
  {Fan}}, \bibinfo {author} {\bibfnamefont {H.}~\bibnamefont {Celik}}, \bibinfo
  {author} {\bibfnamefont {J.}~\bibnamefont {Wu}}, \bibinfo {author}
  {\bibfnamefont {C.}~\bibnamefont {Ni}}, \bibinfo {author} {\bibfnamefont
  {K.-J.}\ \bibnamefont {Lee}}, \bibinfo {author} {\bibfnamefont {V.~O.}\
  \bibnamefont {Lorenz}}, \ and\ \bibinfo {author} {\bibfnamefont {J.~Q.}\
  \bibnamefont {Xiao}},\ }\href {\doibase 10.1038/ncomms4042} {\bibfield
  {journal} {\bibinfo  {journal} {Nature Communications}\ }\textbf {\bibinfo
  {volume} {5}},\ \bibinfo {pages} {3042} (\bibinfo {year} {2014})}\BibitemShut
  {NoStop}%
\bibitem [{\citenamefont {Pai}\ \emph {et~al.}(2015)\citenamefont {Pai},
  \citenamefont {Ou}, \citenamefont {Vilela-Leao}, \citenamefont {Ralph},\ and\
  \citenamefont {Buhrman}}]{Pai2015}%
  \BibitemOpen
  \bibfield  {author} {\bibinfo {author} {\bibfnamefont {C.~F.}\ \bibnamefont
  {Pai}}, \bibinfo {author} {\bibfnamefont {Y.}~\bibnamefont {Ou}}, \bibinfo
  {author} {\bibfnamefont {L.~H.}\ \bibnamefont {Vilela-Leao}}, \bibinfo
  {author} {\bibfnamefont {D.~C.}\ \bibnamefont {Ralph}}, \ and\ \bibinfo
  {author} {\bibfnamefont {R.~A.}\ \bibnamefont {Buhrman}},\ }\href {\doibase
  10.1103/PhysRevB.92.064426} {\bibfield  {journal} {\bibinfo  {journal}
  {Physical Review B}\ }\textbf {\bibinfo {volume} {92}},\ \bibinfo {pages}
  {064426} (\bibinfo {year} {2015})}\BibitemShut {NoStop}%
\bibitem [{\citenamefont {Skinner}\ \emph {et~al.}(2014)\citenamefont
  {Skinner}, \citenamefont {Wang}, \citenamefont {Hindmarch}, \citenamefont
  {Rushforth}, \citenamefont {Irvine}, \citenamefont {Heiss}, \citenamefont
  {Kurebayashi},\ and\ \citenamefont {Ferguson}}]{Skinner2014}%
  \BibitemOpen
  \bibfield  {author} {\bibinfo {author} {\bibfnamefont {T.~D.}\ \bibnamefont
  {Skinner}}, \bibinfo {author} {\bibfnamefont {M.}~\bibnamefont {Wang}},
  \bibinfo {author} {\bibfnamefont {a.~T.}\ \bibnamefont {Hindmarch}}, \bibinfo
  {author} {\bibfnamefont {a.~W.}\ \bibnamefont {Rushforth}}, \bibinfo {author}
  {\bibfnamefont {A.}~\bibnamefont {Irvine}}, \bibinfo {author} {\bibfnamefont
  {D.}~\bibnamefont {Heiss}}, \bibinfo {author} {\bibfnamefont
  {H.}~\bibnamefont {Kurebayashi}}, \ and\ \bibinfo {author} {\bibfnamefont
  {a.~J.}\ \bibnamefont {Ferguson}},\ }\href {\doibase 10.1063/1.4864399}
  {\bibfield  {journal} {\bibinfo  {journal} {Applied Physics Letters}\
  }\textbf {\bibinfo {volume} {104}},\ \bibinfo {pages} {062401} (\bibinfo
  {year} {2014})}\BibitemShut {NoStop}%
\bibitem [{\citenamefont {Nguyen}\ \emph {et~al.}(2016)\citenamefont {Nguyen},
  \citenamefont {Ralph},\ and\ \citenamefont {Buhrman}}]{Nguyen2016}%
  \BibitemOpen
  \bibfield  {author} {\bibinfo {author} {\bibfnamefont {M.~H.}\ \bibnamefont
  {Nguyen}}, \bibinfo {author} {\bibfnamefont {D.~C.}\ \bibnamefont {Ralph}}, \
  and\ \bibinfo {author} {\bibfnamefont {R.~A.}\ \bibnamefont {Buhrman}},\
  }\href {\doibase 10.1103/PhysRevLett.116.126601} {\bibfield  {journal}
  {\bibinfo  {journal} {Physical Review Letters}\ }\textbf {\bibinfo {volume}
  {116}},\ \bibinfo {pages} {126601} (\bibinfo {year} {2016})}\BibitemShut
  {NoStop}%
\bibitem [{\citenamefont {Manchon}(2012)}]{Manchon2012}%
  \BibitemOpen
  \bibfield  {author} {\bibinfo {author} {\bibfnamefont {A.}~\bibnamefont
  {Manchon}},\ }\href@noop {} {\bibfield  {journal} {\bibinfo  {journal} {arXiv
  preprint arXiv:1204.4869v1}\ } (\bibinfo {year} {2012})}\BibitemShut
  {NoStop}%
\bibitem [{\citenamefont {Amin}\ and\ \citenamefont
  {Stiles}(2016)}]{Amin2016b}%
  \BibitemOpen
  \bibfield  {author} {\bibinfo {author} {\bibfnamefont {V.~P.}\ \bibnamefont
  {Amin}}\ and\ \bibinfo {author} {\bibfnamefont {M.~D.}\ \bibnamefont
  {Stiles}},\ }\href {\doibase 10.1103/PhysRevB.94.104419} {\bibfield
  {journal} {\bibinfo  {journal} {Physical Review B}\ }\textbf {\bibinfo
  {volume} {94}},\ \bibinfo {pages} {104420} (\bibinfo {year}
  {2016})}\BibitemShut {NoStop}%
\bibitem [{\citenamefont {Fischer}\ \emph {et~al.}(2016)\citenamefont
  {Fischer}, \citenamefont {Vaezi}, \citenamefont {Manchon},\ and\
  \citenamefont {Kim}}]{Fischer2016}%
  \BibitemOpen
  \bibfield  {author} {\bibinfo {author} {\bibfnamefont {M.~H.}\ \bibnamefont
  {Fischer}}, \bibinfo {author} {\bibfnamefont {A.}~\bibnamefont {Vaezi}},
  \bibinfo {author} {\bibfnamefont {A.}~\bibnamefont {Manchon}}, \ and\
  \bibinfo {author} {\bibfnamefont {E.-a.}\ \bibnamefont {Kim}},\ }\href
  {\doibase 10.1103/PhysRevB.93.125303} {\bibfield  {journal} {\bibinfo
  {journal} {Physical Review B}\ }\textbf {\bibinfo {volume} {93}},\ \bibinfo
  {pages} {125303} (\bibinfo {year} {2016})}\BibitemShut {NoStop}%
\bibitem [{\citenamefont {Shchelushkin}\ and\ \citenamefont
  {Brataas}(2005)}]{Shchelushkin2005b}%
  \BibitemOpen
  \bibfield  {author} {\bibinfo {author} {\bibfnamefont {R.}~\bibnamefont
  {Shchelushkin}}\ and\ \bibinfo {author} {\bibfnamefont {A.}~\bibnamefont
  {Brataas}},\ }\href {\doibase 10.1103/PhysRevB.71.045123} {\bibfield
  {journal} {\bibinfo  {journal} {Physical Review B}\ }\textbf {\bibinfo
  {volume} {71}},\ \bibinfo {pages} {045123} (\bibinfo {year}
  {2005})}\BibitemShut {NoStop}%
\bibitem [{\citenamefont {Hayashi}\ \emph {et~al.}(2014)\citenamefont
  {Hayashi}, \citenamefont {Kim}, \citenamefont {Yamanouchi},\ and\
  \citenamefont {Ohno}}]{Hayashi2014}%
  \BibitemOpen
  \bibfield  {author} {\bibinfo {author} {\bibfnamefont {M.}~\bibnamefont
  {Hayashi}}, \bibinfo {author} {\bibfnamefont {J.}~\bibnamefont {Kim}},
  \bibinfo {author} {\bibfnamefont {M.}~\bibnamefont {Yamanouchi}}, \ and\
  \bibinfo {author} {\bibfnamefont {H.}~\bibnamefont {Ohno}},\ }\href {\doibase
  10.1103/PhysRevB.89.144425} {\bibfield  {journal} {\bibinfo  {journal}
  {Physical Review B}\ }\textbf {\bibinfo {volume} {89}},\ \bibinfo {pages}
  {144425} (\bibinfo {year} {2014})}\BibitemShut {NoStop}%
\bibitem [{\citenamefont {Akyol}\ \emph {et~al.}(2016)\citenamefont {Akyol},
  \citenamefont {Jiang}, \citenamefont {Yu}, \citenamefont {Fan}, \citenamefont
  {Gunes}, \citenamefont {Ekicibil}, \citenamefont {{Khalili Amiri}},\ and\
  \citenamefont {Wang}}]{Akyol2016}%
  \BibitemOpen
  \bibfield  {author} {\bibinfo {author} {\bibfnamefont {M.}~\bibnamefont
  {Akyol}}, \bibinfo {author} {\bibfnamefont {W.}~\bibnamefont {Jiang}},
  \bibinfo {author} {\bibfnamefont {G.}~\bibnamefont {Yu}}, \bibinfo {author}
  {\bibfnamefont {Y.}~\bibnamefont {Fan}}, \bibinfo {author} {\bibfnamefont
  {M.}~\bibnamefont {Gunes}}, \bibinfo {author} {\bibfnamefont
  {A.}~\bibnamefont {Ekicibil}}, \bibinfo {author} {\bibfnamefont
  {P.}~\bibnamefont {{Khalili Amiri}}}, \ and\ \bibinfo {author} {\bibfnamefont
  {K.~L.}\ \bibnamefont {Wang}},\ }\href {\doibase 10.1063/1.4958295}
  {\bibfield  {journal} {\bibinfo  {journal} {Applied Physics Letters}\
  }\textbf {\bibinfo {volume} {109}},\ \bibinfo {pages} {022403} (\bibinfo
  {year} {2016})}\BibitemShut {NoStop}%
\bibitem [{\citenamefont {Ramaswamy}\ \emph {et~al.}(2016)\citenamefont
  {Ramaswamy}, \citenamefont {Qiu}, \citenamefont {Dutta}, \citenamefont
  {Pollard},\ and\ \citenamefont {Yang}}]{Ramaswamy2016}%
  \BibitemOpen
  \bibfield  {author} {\bibinfo {author} {\bibfnamefont {R.}~\bibnamefont
  {Ramaswamy}}, \bibinfo {author} {\bibfnamefont {X.}~\bibnamefont {Qiu}},
  \bibinfo {author} {\bibfnamefont {T.}~\bibnamefont {Dutta}}, \bibinfo
  {author} {\bibfnamefont {S.~D.}\ \bibnamefont {Pollard}}, \ and\ \bibinfo
  {author} {\bibfnamefont {H.}~\bibnamefont {Yang}},\ }\href {\doibase
  10.1063/1.4951674} {\bibfield  {journal} {\bibinfo  {journal} {Applied
  Physics Letters}\ }\textbf {\bibinfo {volume} {108}},\ \bibinfo {pages}
  {202406} (\bibinfo {year} {2016})}\BibitemShut {NoStop}%
\bibitem [{\citenamefont {Sondheimer}(2001)}]{Sondheimer2001}%
  \BibitemOpen
  \bibfield  {author} {\bibinfo {author} {\bibfnamefont {E.}~\bibnamefont
  {Sondheimer}},\ }\href {\doibase 10.1080/00018735200101151} {\bibfield
  {journal} {\bibinfo  {journal} {Advances in Physics}\ }\textbf {\bibinfo
  {volume} {50}},\ \bibinfo {pages} {499} (\bibinfo {year} {2001})}\BibitemShut
  {NoStop}%
\bibitem [{\citenamefont {Kurebayashi}\ \emph {et~al.}(2014)\citenamefont
  {Kurebayashi}, \citenamefont {Sinova}, \citenamefont {Fang}, \citenamefont
  {Irvine}, \citenamefont {Skinner}, \citenamefont {Wunderlich}, \citenamefont
  {Nov{\'{a}}k}, \citenamefont {Campion}, \citenamefont {Gallagher},
  \citenamefont {Vehstedt}, \citenamefont {Z{\^{a}}rbo}, \citenamefont
  {V{\'{y}}born{\'{y}}}, \citenamefont {Ferguson},\ and\ \citenamefont
  {Jungwirth}}]{Kurebayashi2014}%
  \BibitemOpen
  \bibfield  {author} {\bibinfo {author} {\bibfnamefont {H.}~\bibnamefont
  {Kurebayashi}}, \bibinfo {author} {\bibfnamefont {J.}~\bibnamefont {Sinova}},
  \bibinfo {author} {\bibfnamefont {D.}~\bibnamefont {Fang}}, \bibinfo {author}
  {\bibfnamefont {A.}~\bibnamefont {Irvine}}, \bibinfo {author} {\bibfnamefont
  {T.~D.}\ \bibnamefont {Skinner}}, \bibinfo {author} {\bibfnamefont
  {J.}~\bibnamefont {Wunderlich}}, \bibinfo {author} {\bibfnamefont
  {V.}~\bibnamefont {Nov{\'{a}}k}}, \bibinfo {author} {\bibfnamefont {R.~P.}\
  \bibnamefont {Campion}}, \bibinfo {author} {\bibfnamefont {B.~L.}\
  \bibnamefont {Gallagher}}, \bibinfo {author} {\bibfnamefont {E.~K.}\
  \bibnamefont {Vehstedt}}, \bibinfo {author} {\bibfnamefont {L.~P.}\
  \bibnamefont {Z{\^{a}}rbo}}, \bibinfo {author} {\bibfnamefont
  {K.}~\bibnamefont {V{\'{y}}born{\'{y}}}}, \bibinfo {author} {\bibfnamefont
  {A.~J.}\ \bibnamefont {Ferguson}}, \ and\ \bibinfo {author} {\bibfnamefont
  {T.}~\bibnamefont {Jungwirth}},\ }\href {\doibase 10.1038/nnano.2014.15}
  {\bibfield  {journal} {\bibinfo  {journal} {Nature Nanotechnology}\ }\textbf
  {\bibinfo {volume} {9}},\ \bibinfo {pages} {211} (\bibinfo {year}
  {2014})}\BibitemShut {NoStop}%
\bibitem [{\citenamefont {Zwierzycki}\ \emph {et~al.}(2005)\citenamefont
  {Zwierzycki}, \citenamefont {Tserkovnyak}, \citenamefont {Kelly},
  \citenamefont {Brataas},\ and\ \citenamefont {Bauer}}]{Zwierzycki2005}%
  \BibitemOpen
  \bibfield  {author} {\bibinfo {author} {\bibfnamefont {M.}~\bibnamefont
  {Zwierzycki}}, \bibinfo {author} {\bibfnamefont {Y.}~\bibnamefont
  {Tserkovnyak}}, \bibinfo {author} {\bibfnamefont {P.~J.}\ \bibnamefont
  {Kelly}}, \bibinfo {author} {\bibfnamefont {A.}~\bibnamefont {Brataas}}, \
  and\ \bibinfo {author} {\bibfnamefont {G.~E.~W.}\ \bibnamefont {Bauer}},\
  }\href {\doibase 10.1103/PhysRevB.71.064420} {\bibfield  {journal} {\bibinfo
  {journal} {Phys. Rev. B}\ }\textbf {\bibinfo {volume} {71}},\ \bibinfo
  {pages} {064420} (\bibinfo {year} {2005})}\BibitemShut {NoStop}%
\bibitem [{\citenamefont {Qiu}\ \emph {et~al.}(2016)\citenamefont {Qiu},
  \citenamefont {Legrand}, \citenamefont {He}, \citenamefont {Wu},
  \citenamefont {Yu}, \citenamefont {Ramaswamy}, \citenamefont {Manchon},\ and\
  \citenamefont {Yang}}]{Qiu2016}%
  \BibitemOpen
  \bibfield  {author} {\bibinfo {author} {\bibfnamefont {X.}~\bibnamefont
  {Qiu}}, \bibinfo {author} {\bibfnamefont {W.}~\bibnamefont {Legrand}},
  \bibinfo {author} {\bibfnamefont {P.}~\bibnamefont {He}}, \bibinfo {author}
  {\bibfnamefont {Y.}~\bibnamefont {Wu}}, \bibinfo {author} {\bibfnamefont
  {J.}~\bibnamefont {Yu}}, \bibinfo {author} {\bibfnamefont {R.}~\bibnamefont
  {Ramaswamy}}, \bibinfo {author} {\bibfnamefont {A.}~\bibnamefont {Manchon}},
  \ and\ \bibinfo {author} {\bibfnamefont {H.}~\bibnamefont {Yang}},\ }\href
  {\doibase 10.1103/PhysRevLett.117.217206} {\bibfield  {journal} {\bibinfo
  {journal} {Physical Review Letters}\ }\textbf {\bibinfo {volume} {117}},\
  \bibinfo {pages} {217206} (\bibinfo {year} {2016})}\BibitemShut {NoStop}%
\bibitem [{\citenamefont {Lee}\ \emph {et~al.}(2015)\citenamefont {Lee},
  \citenamefont {Go}, \citenamefont {Manchon}, \citenamefont {Haney},
  \citenamefont {Stiles}, \citenamefont {Lee},\ and\ \citenamefont
  {Lee}}]{Lee2015}%
  \BibitemOpen
  \bibfield  {author} {\bibinfo {author} {\bibfnamefont {K.-S.}\ \bibnamefont
  {Lee}}, \bibinfo {author} {\bibfnamefont {D.}~\bibnamefont {Go}}, \bibinfo
  {author} {\bibfnamefont {A.}~\bibnamefont {Manchon}}, \bibinfo {author}
  {\bibfnamefont {P.~M.}\ \bibnamefont {Haney}}, \bibinfo {author}
  {\bibfnamefont {M.~D.}\ \bibnamefont {Stiles}}, \bibinfo {author}
  {\bibfnamefont {H.-W.}\ \bibnamefont {Lee}}, \ and\ \bibinfo {author}
  {\bibfnamefont {K.-J.}\ \bibnamefont {Lee}},\ }\href {\doibase
  10.1103/PhysRevB.91.144401} {\bibfield  {journal} {\bibinfo  {journal}
  {Physical Review B}\ }\textbf {\bibinfo {volume} {91}},\ \bibinfo {pages}
  {144401} (\bibinfo {year} {2015})}\BibitemShut {NoStop}%
\bibitem [{\citenamefont {Pauyac}\ \emph {et~al.}(2013)\citenamefont {Pauyac},
  \citenamefont {Wang}, \citenamefont {Chshiev},\ and\ \citenamefont
  {Manchon}}]{Pauyac2013}%
  \BibitemOpen
  \bibfield  {author} {\bibinfo {author} {\bibfnamefont {C.~O.}\ \bibnamefont
  {Pauyac}}, \bibinfo {author} {\bibfnamefont {X.}~\bibnamefont {Wang}},
  \bibinfo {author} {\bibfnamefont {M.}~\bibnamefont {Chshiev}}, \ and\
  \bibinfo {author} {\bibfnamefont {A.}~\bibnamefont {Manchon}},\ }\href@noop
  {} {\bibfield  {journal} {\bibinfo  {journal} {Applied Physics Letters}\
  }\textbf {\bibinfo {volume} {102}},\ \bibinfo {pages} {252403} (\bibinfo
  {year} {2013})}\BibitemShut {NoStop}%
\bibitem [{\citenamefont {Hals}\ and\ \citenamefont
  {Brataas}(2014)}]{Hals2014}%
  \BibitemOpen
  \bibfield  {author} {\bibinfo {author} {\bibfnamefont {K.}~\bibnamefont
  {Hals}}\ and\ \bibinfo {author} {\bibfnamefont {A.}~\bibnamefont {Brataas}},\
  }\href {\doibase 10.1103/PhysRevB.89.064426} {\bibfield  {journal} {\bibinfo
  {journal} {Physical Review B}\ }\textbf {\bibinfo {volume} {89}},\ \bibinfo
  {pages} {064426} (\bibinfo {year} {2014})}\BibitemShut {NoStop}%
\bibitem [{\citenamefont {{\v{Z}}elezn{\'{y}}}\ \emph
  {et~al.}(2017)\citenamefont {{\v{Z}}elezn{\'{y}}}, \citenamefont {Gao},
  \citenamefont {Manchon}, \citenamefont {Freimuth}, \citenamefont {Mokrousov},
  \citenamefont {Zemen}, \citenamefont {Ma{\v{s}}ek}, \citenamefont {Sinova},\
  and\ \citenamefont {Jungwirth}}]{Zelezny2017}%
  \BibitemOpen
  \bibfield  {author} {\bibinfo {author} {\bibfnamefont {J.}~\bibnamefont
  {{\v{Z}}elezn{\'{y}}}}, \bibinfo {author} {\bibfnamefont {H.}~\bibnamefont
  {Gao}}, \bibinfo {author} {\bibfnamefont {A.}~\bibnamefont {Manchon}},
  \bibinfo {author} {\bibfnamefont {F.}~\bibnamefont {Freimuth}}, \bibinfo
  {author} {\bibfnamefont {Y.}~\bibnamefont {Mokrousov}}, \bibinfo {author}
  {\bibfnamefont {J.}~\bibnamefont {Zemen}}, \bibinfo {author} {\bibfnamefont
  {J.}~\bibnamefont {Ma{\v{s}}ek}}, \bibinfo {author} {\bibfnamefont
  {J.}~\bibnamefont {Sinova}}, \ and\ \bibinfo {author} {\bibfnamefont
  {T.}~\bibnamefont {Jungwirth}},\ }\href {\doibase 10.1103/PhysRevB.95.014403}
  {\bibfield  {journal} {\bibinfo  {journal} {Physical Review B}\ }\textbf
  {\bibinfo {volume} {95}},\ \bibinfo {pages} {014403} (\bibinfo {year}
  {2017})}\BibitemShut {NoStop}%
\end{thebibliography}%

\end{document}